\documentclass[twocolumn, times]{aastex63}

\usepackage{rotating}
\usepackage{booktabs}
\usepackage[version=4]{mhchem}
\usepackage{multirow}
\usepackage{longtable}
\usepackage{graphicx}
\usepackage{color}
\usepackage{xcolor}

\newcommand{\tr}[1]{#1}

\usepackage{comment}
\begin{document}
\title{\large{\texttt{NEXONN}: Advancing the \texttt{NEXOTRANS} Exoplanet Atmospheric Retrieval Framework with Transformer-Based Neural Posterior Estimation for JWST Transmission Spectroscopy of WASP-39b}}

\correspondingauthor{Liton Majumdar}
\email{liton@niser.ac.in, dr.liton.majumdar@gmail.com}

\author[0009-0001-3342-3996]{Tasneem Basra Khan}
\affiliation{Exoplanets and Planetary Formation Group, School of Earth and Planetary Sciences, National Institute of Science Education and Research, Jatni 752050, Odisha, India}
\affiliation{Homi Bhabha National Institute, Training School Complex, Anushaktinagar, Mumbai 400094, India}

\author[0000-0001-7031-8039]{Liton Majumdar}
\affiliation{Exoplanets and Planetary Formation Group, School of Earth and Planetary Sciences, National Institute of Science Education and Research, Jatni 752050, Odisha, India}
\affiliation{Homi Bhabha National Institute, Training School Complex, Anushaktinagar, Mumbai 400094, India}

\author[0009-0002-6571-9406]{Mithun Patra}
\affiliation{Exoplanets and Planetary Formation Group, School of Earth and Planetary Sciences, National Institute of Science Education and Research, Jatni 752050, Odisha, India}
\affiliation{Homi Bhabha National Institute, Training School Complex, Anushaktinagar, Mumbai 400094, India}

\author[0009-0007-7880-0250]{Tonmoy Deka}
\affiliation{Exoplanets and Planetary Formation Group, School of Earth and Planetary Sciences, National Institute of Science Education and Research, Jatni 752050, Odisha, India}
\affiliation{Homi Bhabha National Institute, Training School Complex, Anushaktinagar, Mumbai 400094, India}

\author[0009-0002-6571-9406]{Swastik Dewan}
\affiliation{Exoplanets and Planetary Formation Group, School of Earth and Planetary Sciences, National Institute of Science Education and Research, Jatni 752050, Odisha, India}
\affiliation{Homi Bhabha National Institute, Training School Complex, Anushaktinagar, Mumbai 400094, India}

\author[ 0009-0002-4995-9346]{Priyankush Ghosh}
\affiliation{Exoplanets and Planetary Formation Group, School of Earth and Planetary Sciences, National Institute of Science Education and Research, Jatni 752050, Odisha, India}
\affiliation{Homi Bhabha National Institute, Training School Complex, Anushaktinagar, Mumbai 400094, India}

\author [0009-0006-6349-1451] {Vishal Meena}
\affiliation{Exoplanets and Planetary Formation Group, School of Earth and Planetary Sciences, National Institute of Science Education and Research, Jatni 752050, Odisha, India}
\affiliation{Homi Bhabha National Institute, Training School Complex, Anushaktinagar, Mumbai 400094, India}


\begin{abstract}

The James Webb Space Telescope is delivering unprecedented constraints on exoplanet atmospheres, but computationally intensive Bayesian retrievals challenge the transition from individual planets to population-level studies of planetary formation and evolution. We introduce \texttt{NEXONN}, a transformer-conditioned normalizing-flow framework within \texttt{NEXOTRANS} that enables rapid, amortized atmospheric posterior estimation with uncertainty quantification. The model learns spectral dependencies and atmospheric parameter correlations while incorporating \texttt{PandExo}-based, wavelength-dependent JWST noise during training. We benchmark \texttt{NEXONN} against \texttt{PyMultiNest} using simulated transmission spectra with known atmospheric parameters and JWST observations of WASP-39~b spanning $0.6$--$12\,\mu\mathrm{m}$. For the simulated MIRI/LRS spectrum, the eight marginalized posteriors achieve a mean overlap of $80.32\%$, demonstrating substantial agreement in the inferred atmospheric constraints and uncertainties. Retrievals of observed MIRI/LRS and combined NIRISS, NIRSpec, and MIRI spectra recover broadly consistent atmospheric properties. For the full JWST dataset, \texttt{NEXONN} infers C/O = $0.82^{+0.01}_{-0.02}$ and metallicity $\mathrm{[M/H]}=1.13^{+0.04}_{-0.04}$, compared with \texttt{PyMultiNest} values of $0.80^{+0.03}_{-0.01}$ and $1.19^{+0.05}_{-0.03}$, respectively. KL and JS divergences, standardized Wasserstein distances, posterior overlaps, and simulation-based coverage tests assess posterior fidelity and uncertainty calibration, revealing parameter-dependent differences alongside consistent atmospheric interpretations. Once trained, \texttt{NEXONN} performs posterior inference within seconds, offering substantial computational efficiency compared with conventional Bayesian retrieval methods and enabling scalable atmospheric characterization of large exoplanet samples. This framework opens new opportunities for statistically robust population studies and comparative investigations of exoplanet atmospheric composition, formation, and evolution.

\end{abstract}

\keywords{Exoplanets (498); Exoplanet atmospheres (487); Hot Jupiters (753); Extrasolar gaseous giant planets (509); Exoplanet atmospheric structure (2310); Exoplanet atmospheric composition (2021)}

\section{\textbf{Introduction}} \label{sec:intro}

 The discovery of more than 6000 confirmed exoplanets\footnote{\url{https://exoplanetarchive.ipac.caltech.edu/}}, enabled by major space-based surveys such as the Kepler Space Telescope and the Transiting Exoplanet Survey Satellite (TESS), has revealed a remarkable diversity of planetary systems \citep[see, e.g.,][for reviews]{Seager_2010,Cloutier2026}. These worlds span a broad parameter space in mass, radius, equilibrium temperature, orbital architecture, and host-star properties, encompassing planetary classes from ultra-hot Jupiters to temperate sub-Neptunes and rocky super-Earths \citep[see, e.g.,][for reviews]{Madhusudhan_2019,fortney2024characterizing,Manoj25,Valencia25}. The growing diversity of known exoplanets has increasingly motivated efforts to understand the physical and chemical processes that shape their formation and evolution. Atmospheric characterization is central to these efforts, providing an observational pathway to constrain planetary composition, thermal structure, and evolutionary history. For example, the atmospheric \tr{carbon-to-oxygen ratio (C/O)}  can provide insights into a planet's formation environment relative to major ice lines in the protoplanetary disk and its subsequent migration history \citep{Madhusudhan_2012,Dash22,He26}. Atmospheric composition also reflects the interplay of thermochemical equilibrium, disequilibrium chemistry, and photochemical processes \citep{Tsai_2023,Ghosh26a}, while, for terrestrial planets, it may additionally carry signatures of surface and interior processes \citep{lichtenberg2024super,sahu2025unveiling}. Inferring these physical and chemical properties from observations requires atmospheric retrieval techniques that couple spectroscopic data with radiative-transfer models \citep{Madhu18}. Transmission spectroscopy probes the atmospheric limb during primary transit, enabling constraints on molecular abundances and cloud and haze properties \citep{Sing_2015}, whereas secondary-eclipse emission spectroscopy probes the planetary dayside, providing constraints on atmospheric composition and the pressure--temperature structure, including features such as thermal inversions \citep{Line_2013}. These observational techniques therefore form the basis for translating exoplanet spectra into constraints on atmospheric physics and chemistry.


\tr{While early observations with the Hubble Space Telescope (HST) and
Spitzer Space Telescope laid the foundation for exoplanet atmospheric
characterization (see \citet{Madhusudhan_2019} for a comprehensive
review), the advent of the James Webb Space Telescope (JWST) has
substantially advanced the field through its unprecedented sensitivity
and broad wavelength coverage, extending from approximately
$0.6$ to $28\,\mu\mathrm{m}$ \citep{Gardner_2023}.
JWST observations now provide increasingly detailed constraints on
atmospheric composition, pressure--temperature structures, clouds
and aerosols, elemental abundance ratios, and disequilibrium chemistry
across a diverse range of planets, from hot Jupiters to sub-Neptunes
and super-Earths
\citep[see, e.g.,][for recent reviews]{espinoza2025highlights,Hinkley26}.
The growing JWST archive, together with future observations from
the Atmospheric Remote-sensing Infrared Exoplanet Large-survey (ARIEL)
mission, will facilitate comparative atmospheric studies across
increasingly diverse planetary samples \citep{Changeat25}.
These advances highlight the need for accurate and computationally
scalable atmospheric retrieval frameworks.}

\tr{Atmospheric retrievals combine forward models of planetary spectra
\citep{Deka_2025,MacDonald_2023,Molliere_2019,Taurex_2021,vira_2024}
with Bayesian inference, commonly employing Markov Chain Monte Carlo
(MCMC) or nested-sampling methods
\citep{Trotta_2008,Goodman_2010,Hastings_1970,Skilling;2006,Higson_2019}.
However, conventional Bayesian retrievals often require numerous
forward-model evaluations and substantial computational resources,
making their repeated application to large planetary samples
challenging. This motivates faster inference methods that preserve
access to posterior probability distributions.}

\tr{Machine learning (ML) offers a promising approach to accelerating
atmospheric retrievals, with simulation-based inference (SBI)
providing a framework for learning statistical relationships
between simulated data and atmospheric parameters
\citep{cranmer_2020}.
Previous ML-based approaches to exoplanet atmospheric retrieval
include random forests \citep{Hela_2018,RF_2020},
deep neural networks
\citep{Exogan_2018,Inara_2018,Cob_2019,exocnn_2022},
and surrogate models
\citep{himes_2022,unlu2023reproducingbayesianposteriordistributions}.
More recently, neural posterior estimation (NPE), including
approaches based on normalizing flow (NF) models, has gained
increasing attention as a strategy for directly approximating
atmospheric posterior distributions \citep{Vasist_2023,Yip_2024,flopity_2024,Gebhard_2025,
faster_2025,Deka_2025}.
Unlike conventional Bayesian sampling methods, amortized NPE
learns conditional posterior distributions from simulated spectra,
enabling rapid inference for new observations after training.
However, its reliability depends on how well the training simulations
represent the atmospheric parameter space, forward-model physics,
and observational noise characteristics.
Robust and scalable SBI-based retrievals therefore require
diverse training spectra and realistic instrumental uncertainties.}

Motivated by these requirements, we introduce the \texttt{NEXOTRANS}
Neural Network (\texttt{NEXONN}), an SBI framework that combines
\tr{transformer-conditioned normalizing flows (TcNF)}
\citep{tnaf_2024}, realistic JWST noise simulations, and the
physical flexibility of \texttt{NEXOTRANS}.
The \tr{transformer} uses self-attention to encode correlations
across the input spectrum and conditions the \tr{NF},
which learns the multidimensional posterior distribution of
atmospheric parameters.
\texttt{NEXONN} incorporates \texttt{PandExo}-based noise
realizations during training to represent wavelength-dependent
uncertainties and specific JWST instrumental configurations,
while its integration with \texttt{NEXOTRANS} \citep{Deka_2025}
enables training spectra to span different atmospheric chemistry,
pressure--temperature, and cloud prescriptions.
This represents a substantial extension of the previous ML
implementation in \texttt{NEXOTRANS}, which used a stacking
regressor combining random forests, k-nearest neighbors, and
gradient boosting for parameter estimation.
In contrast, \texttt{NEXONN} performs multidimensional \tr{NPE}
with associated uncertainty quantification.
Thus, the key advance of \texttt{NEXONN} is the integration
of probabilistic \tr{NPE}, instrument-aware JWST noise modeling,
and physically diverse forward modeling within a unified
framework for rapid and scalable atmospheric retrievals.

In this work, we validate \texttt{NEXONN} using both simulated
spectra with known atmospheric parameters and real JWST
observations, benchmarking its posterior distributions against
\texttt{PyMultiNest} retrievals performed within the same
\texttt{NEXOTRANS} framework.
We apply the method to the JWST MIRI/LRS transmission spectrum
of WASP-39~b ($5$--$12\,\mu\mathrm{m}$) and subsequently to its
combined NIRISS/SOSS+NIRSpec/PRISM+MIRI/LRS spectrum spanning
approximately $0.6$--$12\,\mu\mathrm{m}$.
We quantitatively compare the resulting posterior distributions
and evaluate the statistical calibration of the inferred
uncertainties, testing whether amortized neural inference can
recover the principal constraints of conventional Bayesian
retrievals while reducing subsequent inference to timescales
of seconds.
The remainder of this paper is organized as follows.
Sections~\ref{normalizing flows} and~\ref{transformers} describe
the \tr{NF} and \tr{transformer} components of \texttt{NEXONN},
respectively.
Section~\ref{Architecture} presents the architecture, training
procedure, and retrieval setup.
Section~\ref{sec:results} presents the results, while
Section~\ref{sec:discussion} discusses the implications,
limitations, and future development of the framework.

\section{\textbf{\texttt{NEXONN}: A Machine-Learning Retrieval Framework within \texttt{NEXOTRANS}}}
\label{sec:framework}

Atmospheric retrieval is an inverse problem commonly formulated within a Bayesian framework, in which the posterior distribution of the atmospheric parameters, $\theta$, is inferred from an observed spectrum, $\mathbf{x}$, and its associated uncertainties. According to Bayes' theorem,

\begin{equation}
    \mathrm{P}(\theta \mid \mathbf{x}) =
    \frac{\mathrm{P}(\mathbf{x} \mid \theta)\mathrm{P}(\theta)}
    {\mathrm{P}(\mathbf{x})},
    \label{eq:bayes}
\end{equation}
where $\mathrm{P}(\theta \mid \mathbf{x})$ is the posterior distribution of the atmospheric parameters, $\mathrm{P}(\theta)$ is the prior distribution, $\mathrm{P}(\mathbf{x} \mid \theta)$ is the likelihood, and $\mathrm{P}(\mathbf{x})$ is the Bayesian evidence.

The Bayesian evidence in Equation~\ref{eq:bayes} is obtained by marginalizing the likelihood over the prior distribution,

\begin{equation}
    \mathrm{P}(\mathbf{x}) =
    \int
    \mathrm{P}(\mathbf{x} \mid \theta)
    \mathrm{P}(\theta)
    \,\mathrm{d}\theta.
    \label{eq:evidence}
\end{equation}

\tr{Evaluating posterior distributions can become computationally
expensive for high-dimensional atmospheric models, particularly
when the parameter space contains strong degeneracies and
complex posterior structures. As the number of retrieved
parameters increases, conventional Bayesian sampling methods
may require substantially more forward-model evaluations
to adequately explore the parameter space and estimate
posterior distributions and Bayesian evidence.}

\begin{figure*}
    \centering
    \includegraphics[width=\linewidth]{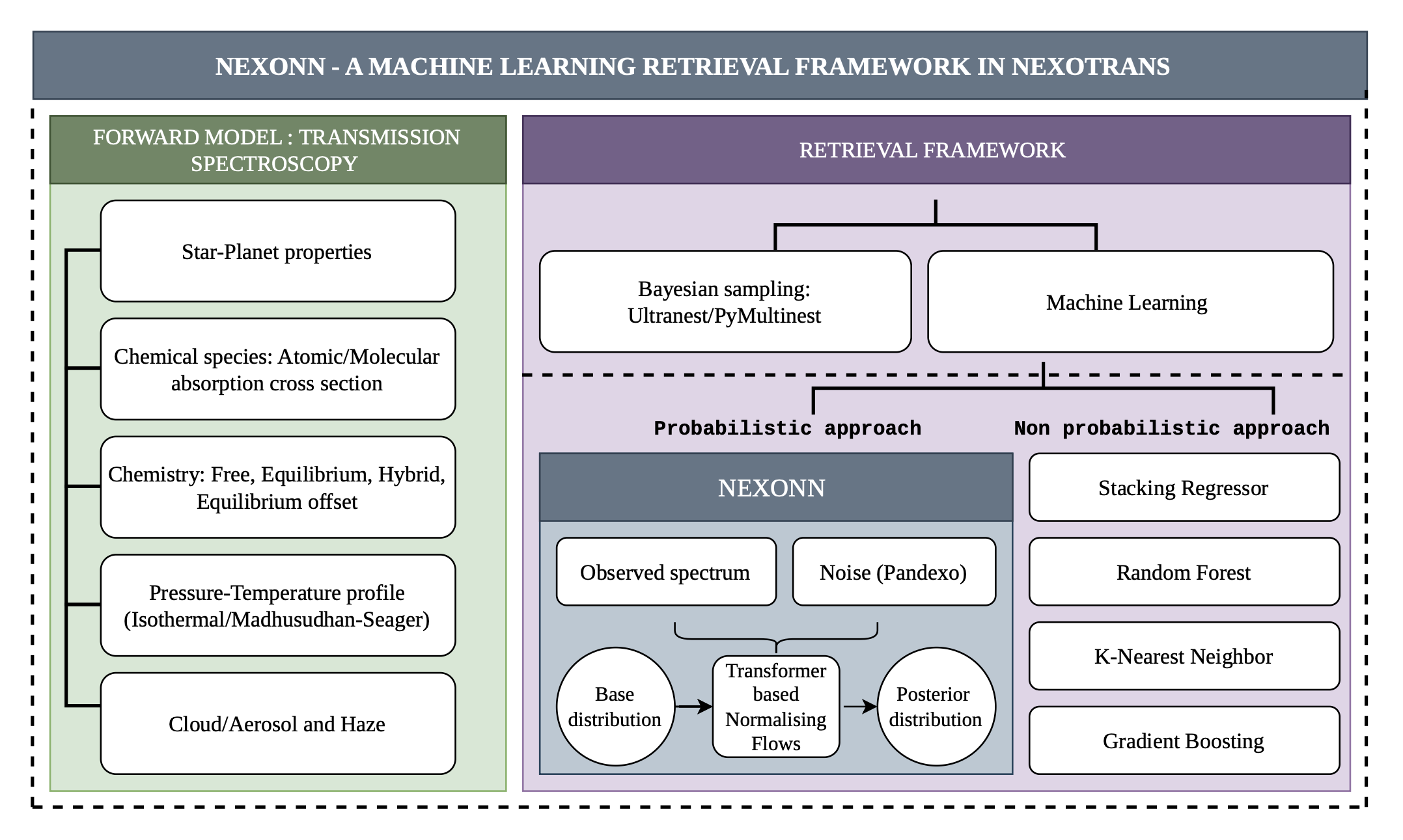}
    \caption{Overview of the \texttt{NEXOTRANS} framework \citep{Deka_2025}, showing the forward-modeling component (green box), existing retrieval methods (purple box), and the \texttt{NEXONN} framework introduced in this work (gray box). The forward model generates transmission spectra by solving the radiative-transfer equation for specified atmospheric properties, including chemical composition, pressure--temperature structure, and cloud prescriptions. The existing retrieval module includes Bayesian sampling methods such as \texttt{PyMultiNest} and \texttt{UltraNest}, together with the machine-learning-based Stacking Regressor, which combines Random Forest, Gradient Boosting, and K-Nearest Neighbors. \texttt{NEXONN} extends this framework using \tr{TcNF} to perform probabilistic posterior estimation of atmospheric parameters.}
    \label{nexonn architecture}
\end{figure*}

Machine-learning approaches provide an alternative strategy by
amortizing the computational cost over a training phase.
Once trained, \tr{NPE} can rapidly infer posterior distributions
for new observations without repeating the sampling procedure
required by conventional Bayesian retrievals.
In this work, we introduce \texttt{NEXONN}, a simulation-based
inference framework within \texttt{NEXOTRANS} \citep{Deka_2025}
that employs \tr{TcNF} for atmospheric posterior estimation.
Figure~\ref{nexonn architecture} illustrates the integration
of \texttt{NEXONN} within the broader \texttt{NEXOTRANS}
framework.
The following subsections describe the theoretical foundations,
architecture, training procedure, and retrieval setup of
\texttt{NEXONN}.

\subsection{\textbf{Normalizing Flows}}
\label{normalizing flows}

NF models \citep{normalising_flows} have emerged as powerful
tools for density estimation within the paradigm of SBI.
NFs model complex probability distributions by transforming
a simple base distribution, typically a Gaussian, through a
sequence of invertible, bijective, and differentiable transformations. This makes them well suited to exoplanet atmospheric retrievals, where the goal is to infer potentially complex and correlated posterior distributions rather than only point estimates. Several recent atmospheric retrieval frameworks have demonstrated the potential of SBI and flow-based approaches for efficient posterior estimation \citep{Vasist_2023,Yip_2024,flopity_2024,Gebhard_2025,faster_2025}. In the context of exoplanet atmospheric retrievals, \citet{Vasist_2023} applied NF within a variational inference framework, while more recent work by \citet{Gebhard_2025} explored continuous normalizing flows (CNF).

For a CNF, the posterior density is governed by the change-of-variables formula,

\begin{equation}
\label{eq:change_of_variable}
    \mathrm{p}(\theta \mid \mathbf{x}) =
    \mathrm{p}_{\mathbf{z}}\!\left(\mathrm{f}(\theta \mid \mathbf{x})\right)
    \left|
    \det\left(
    \frac{\partial \mathrm{f}(\theta \mid \mathbf{x})}{\partial \theta}
    \right)
    \right|,
\end{equation}

where $\theta$ represents the atmospheric parameters, such as temperature, molecular abundances, and aerosol properties, and $\mathbf{x}$ denotes the conditioning variable, here the observed JWST spectrum. The flow maps the target distribution to a simple base distribution, typically a multivariate Gaussian $\mathrm{p}_{\mathbf{z}}$, with $\mathbf{z}\sim\mathcal{N}(0,\mathbf{I})$. Equivalently, the generative direction maps the latent variable $\mathbf{z}$ to the atmospheric parameters through a sequence of invertible transformations,
$\theta = \mathrm{f}_K \circ \mathrm{f}_{K-1} \circ \cdots \circ \mathrm{f}_1(\mathbf{z})$.
Because the transformation is bijective, $\mathbf{z},\theta\in\mathbb{R}^{d}$. The determinant of the Jacobian accounts for the local change in probability density induced by the transformation. The flow is trained by minimizing the negative log-likelihood,

\begin{equation}
\label{eq:loss_function}
    \mathcal{L} =
    -\log \mathrm{p}_{\mathbf{z}}\!\left(\mathrm{f}(\theta \mid \mathbf{x})\right)
    -\log\left|
    \det\left(
    \frac{\partial \mathrm{f}(\theta \mid \mathbf{x})}{\partial \theta}
    \right)
    \right|.
\end{equation}

A major computational consideration in NF models is the efficient
evaluation of the Jacobian determinant. Autoregressive constructions
address this by imposing a triangular Jacobian structure, allowing
the determinant to be evaluated efficiently.
Several NF architectures have been developed, including
inverse autoregressive flows (IAFs) \citep{Kingma_2016},
neural autoregressive flows (NAFs) \citep{Huang_2018},
and block neural autoregressive flows (BNAFs) \citep{bnf_2018}.
In \texttt{NEXONN}, we adopt a transformer-based neural
autoregressive flow (T-NAF), following \citet{tnaf_2024},
in which a transformer attention mechanism
\citep{transformer_2017} provides the conditioning
information for the autoregressive flow.

\subsection{\textbf{Transformer-conditioned Normalizing Flows}}
\label{transformers}

Transformers \citep{transformer_2017} are widely used architectures in natural-language processing, computer vision, and other machine-learning applications. Their key component is the attention mechanism, which enables the model to learn relationships among elements of an input sequence. In a simplified representation, the output at layer $m$ can be written as

\begin{equation}
\label{eq:layer_output}
    \mathbf{Y}^{(m)} =
    \mathbf{X}^{(m-1)}\mathbf{A}^{(m)},
\end{equation}

where $\mathbf{A}^{(m)}$ is the attention matrix, $\mathbf{X}^{(m-1)}$ is the input representation from the preceding layer, and $m$ denotes the transformer layer index. Further mathematical details of the attention mechanism are provided by \citet{Turner_2026}.

\begin{figure}
    \centering
    \includegraphics[width=\linewidth]{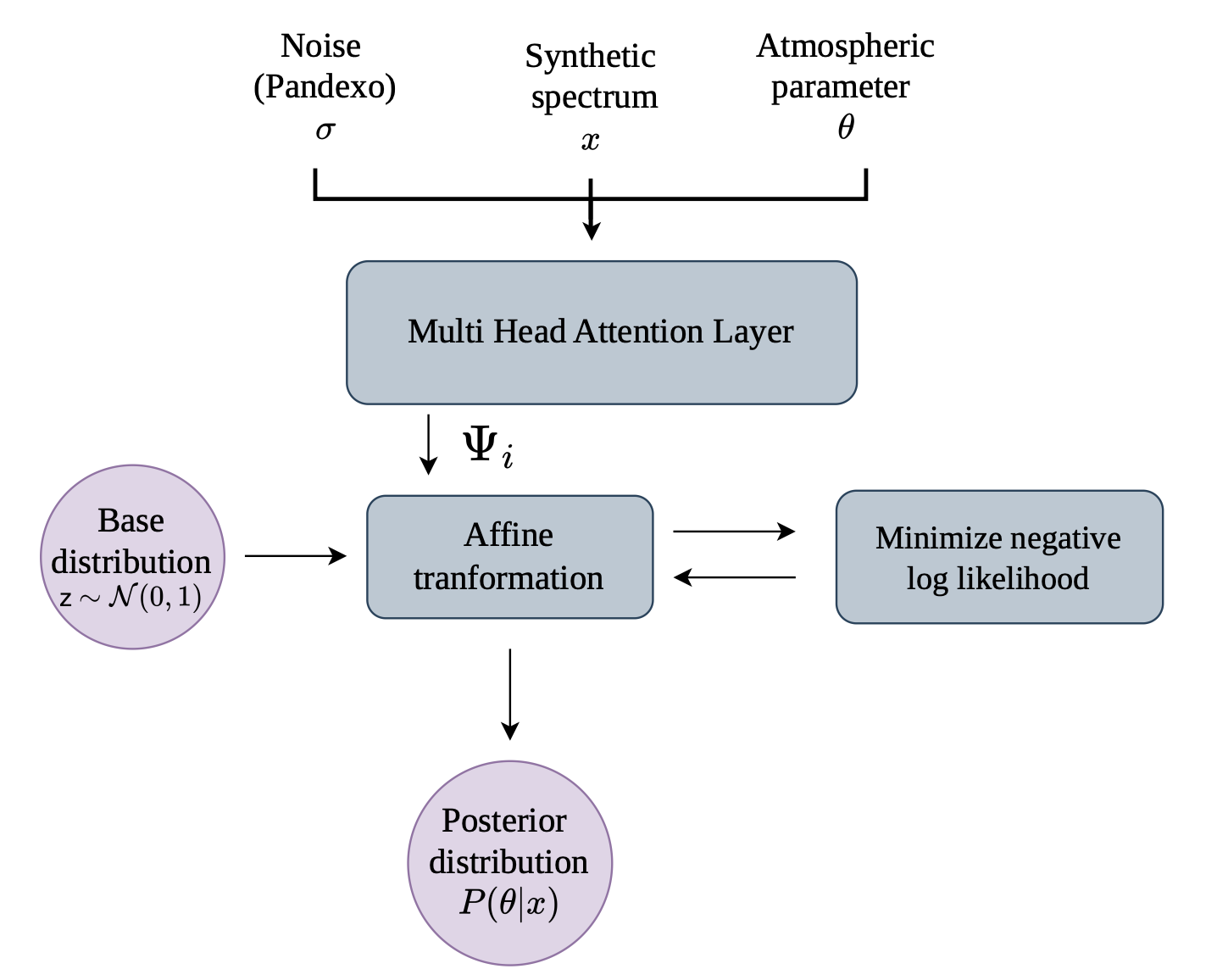}
    \caption{Schematic representation of the \texttt{NEXONN} training framework. Synthetic spectra, $\mathbf{x}$, are generated using \texttt{NEXOTRANS} \citep{Deka_2025}, with \texttt{PandExo}-based noise realizations incorporated during training to represent wavelength-dependent observational uncertainties. The resulting spectra are processed by the \tr{transformer}, which provides the conditioning representation for the \tr{NF}. The flow learns the invertible transformation between the atmospheric parameter space and the latent distribution. Training is performed by minimizing the \tr{NF} negative log-likelihood given in Equation~\ref{eq:loss_function}.}
    \label{training-flow}
\end{figure}

\texttt{NEXONN} combines a \tr{transformer} with an autoregressive \tr{NF} to model the multidimensional posterior distribution of exoplanet atmospheric parameters. The architecture is illustrated in Figure~\ref{training-flow}. The \tr{transformer} receives two distinct sets of tokens. The first represents the spectrum, with each spectral element encoded by three quantities: wavelength ($\lambda$), transit depth ($x$), and its $1\sigma$ uncertainty ($\sigma$). No attention mask is applied to the spectral tokens, allowing each spectral element to attend to all others and enabling the model to capture correlations across the spectrum. The second set represents the atmospheric parameters, $\theta_d$, each encoded as a 128-dimensional embedding. A causal, lower-triangular mask is applied to the parameter tokens so that each parameter can attend only to itself and preceding parameters. The parameter representation then queries the unmasked spectral representation through cross-attention, providing the spectral conditioning required for autoregressive posterior estimation.

The current \texttt{NEXONN} implementation uses six stacked flow layers. Each layer applies an element-wise affine transformation to the atmospheric parameters, with the scale and shift functions conditioned by the transformer. For the $i$th parameter, the forward and inverse affine transformations are

\begin{equation}
\label{eq:affine1}
    z_i =
    \left(\theta_i - t_{i,\theta_{<i}}\right)
    \odot
    \exp\left(-s_{i,\theta_{<i}}\right),
\end{equation}

\begin{equation}
\label{eq:affine2}
    \theta_i =
    z_i \odot
    \exp\left(s_{i,\theta_{<i}}\right)
    + t_{i,\theta_{<i}},
\end{equation}

where $s_{i,\theta_{<i}}$ and $t_{i,\theta_{<i}}$ are the scale and shift functions, respectively, conditioned on the preceding atmospheric parameters $\theta_{<i}$. The atmospheric parameter vector $\theta$ is thus mapped autoregressively to the latent variable $\mathbf{z}$, with both having dimensionality $D$. This autoregressive construction results in a triangular Jacobian. For the transformation in Equation~\ref{eq:affine1}, the Jacobian takes the form

\begin{equation}
\label{eq:jacobian}
    \frac{\partial \mathbf{z}}{\partial \theta} =
    \begin{pmatrix}
        \exp(-s_{1,\theta_{<1}}) & 0 & \cdots & 0 \\
        \frac{\partial z_2}{\partial \theta_1} &
        \exp(-s_{2,\theta_{<2}}) & \cdots & 0 \\
        \vdots & \vdots & \ddots & \vdots \\
        \frac{\partial z_D}{\partial \theta_1} &
        \frac{\partial z_D}{\partial \theta_2} &
        \cdots &
        \exp(-s_{D,\theta_{<D}})
    \end{pmatrix}.
\end{equation}

Consequently, the determinant required for density evaluation is obtained directly from the diagonal elements,

\begin{equation}
\label{eq:det_jacobian}
    \det\left(
    \frac{\partial \mathbf{z}}{\partial \theta}
    \right)
    =
    \prod_{i=1}^{D}
    \exp\left(-s_{i,\theta_{<i}}\right).
\end{equation}

More flexible transformations can be used in place of affine transformations, including cumulative distribution function (CDF)-based transformations \citep{normalising_flows,Huang_2018}, shared-CDF transformations \citep{bnf_2018}, and spline transformations \citep{Durkan_2019}. 

\subsection{\textbf{Architecture, Training Procedure, and Retrieval Setup in \texttt{NEXONN}}}
\label{Architecture}

As discussed in Section~\ref{transformers}, \texttt{NEXONN} is a retrieval framework within \texttt{NEXOTRANS} that employs \tr{TcNF} to model the posterior distributions of atmospheric parameters. Building on this architecture, the following subsections describe the generation of synthetic training spectra, the implementation of observational noise, the training and optimization procedure, and the visualization of posterior predictions with \texttt{NEXONN}.

\subsubsection{\textbf{Synthetic Spectra Generation}}
\label{synthetic spectra generation}

Model-dependent atmospheric characterization requires a forward model to generate synthetic spectra across the relevant physical and chemical parameter space. For this purpose, \texttt{NEXONN} uses the \texttt{NEXOTRANS} forward-modeling framework; a detailed description of its transmission-spectrum calculations is provided by \citet{Deka_2025}. \texttt{NEXOTRANS} incorporates four chemistry prescriptions: \tr{free, equilibrium, hybrid equilibrium, and equilibrium offset}, with equilibrium abundances obtained from the \texttt{NEXOCHEM} grid \citep{Deka_2025}. The framework also includes multiple cloud prescriptions, such as uniform and patchy-gray clouds, and pressure--temperature (P--T) profiles, including isothermal and Madhusudhan--Seager profiles \citep{madhu_pt}. \tr{A broad range of chemical species can be included,
such as $\mathrm{Na}$, $\mathrm{K}$, $\mathrm{H_2O}$,
$\mathrm{CO_2}$, $\mathrm{CO}$, $\mathrm{H_2S}$,
$\mathrm{SO_2}$, $\mathrm{CH_4}$, and $\mathrm{HCN}$,
together with aerosol species such as
$\mathrm{MgSiO_3}$  and $\mathrm{ZnS}$.
The model also includes collision-induced absorption
(CIA) cross sections for $\mathrm{H_2}$--$\mathrm{H_2}$
and $\mathrm{H_2}$--$\mathrm{He}$ pairs, which
contribute to the continuum opacity, as described
in detail by \citet{Deka_2025}.}

Generating robust training data requires sampling a sufficiently broad atmospheric parameter space so that the model can learn the underlying physical and spectral relationships. Synthetic spectra are first computed at high spectral resolution and subsequently binned to the resolution appropriate for the observations. Because real observations contain wavelength-dependent uncertainties in the measured transit depths, realistic noise realizations must also be incorporated into the training spectra. The implementation of observational noise in \texttt{NEXONN} is described in the following subsection.

\subsubsection{\textbf{\texttt{PandExo}-Based Noise Implementation for JWST Spectra}}
\label{PandExo}

To incorporate realistic observational uncertainties into the synthetic spectra, \texttt{NEXONN} uses the \texttt{PandExo} noise simulator \citep{pandexo_2017}, which is specifically designed to simulate observations with JWST instruments. The use of \texttt{PandExo}-based noise allows the training spectra to account for wavelength-dependent instrumental and observational uncertainties. The total variance can be represented schematically as

\begin{equation}
\label{PandExo error}
    \sigma_{\mathrm{total}}^2 =
    \sigma_{\mathrm{photon}}^2 +
    \sigma_{\mathrm{bg}}^2 +
    \sigma_{\mathrm{read}}^2 +
    \sigma_{\mathrm{dark}}^2 +
    \sigma_{\mathrm{sys}}^2,
\end{equation}

where $\sigma_{\mathrm{photon}}^2$ represents the photon-noise contribution associated with the stellar flux, $\sigma_{\mathrm{bg}}^2$ represents background contributions, $\sigma_{\mathrm{read}}^2$ and $\sigma_{\mathrm{dark}}^2$ represent detector read noise and dark-current noise, respectively, and $\sigma_{\mathrm{sys}}^2$ denotes an additional systematic noise-floor contribution when included in the adopted noise model. The resulting noise profile is wavelength dependent and is influenced by both the instrumental response and the spectral energy distribution of the host star.

The uncertainty in the wavelength-dependent transit depth, $\sigma_{x,\lambda}$, can be estimated by propagating the uncertainties associated with the in-transit and out-of-transit flux measurements. Assuming comparable uncertainties for the individual in-transit and out-of-transit measurements, the variance can be written schematically as

\begin{equation}
\label{transit depth error}
    \sigma_{x,\lambda}^2 =
    \left(
    \frac{\sigma_{\mathrm{total},\lambda}}
    {\mathrm{F_{\mathrm{out},\lambda}}}
    \right)^2
    \left(
    \frac{1}{\mathrm{N_{\mathrm{in}}}} +
    \frac{1}{\mathrm{N_{\mathrm{out}}}}
    \right),
\end{equation}

where $\mathrm{F_{\mathrm{out}},\lambda}$ is the out-of-transit stellar flux, and $\mathrm{N_{\mathrm{in}}}$ and $\mathrm{N_{\mathrm{out}}}$ denote the numbers of integrations obtained in and out of transit, respectively. For other observing facilities or missions, alternative instrument-specific noise models can be incorporated into the \texttt{NEXONN} framework.

\subsubsection{\textbf{Training and Inference}}
\label{Training}

For each retrieval experiment, we independently generated a training dataset of two million synthetic spectra using the \texttt{NEXOTRANS} forward model, as described in Section~\ref{synthetic spectra generation}. For the MIRI/LRS benchmark, spectra were generated over the $5$--$12\,\mu\mathrm{m}$ wavelength range using an eight-parameter atmospheric model that includes an isothermal pressure--temperature (P--T) profile, haze, and patchy-gray clouds. For the full JWST retrieval, we considered the combined wavelength coverage of NIRISS/SOSS ($0.6$--$2.8\,\mu\mathrm{m}$), NIRSpec/PRISM ($0.5$--$5\,\mu\mathrm{m}$), and MIRI/LRS ($5$--$12\,\mu\mathrm{m}$). {This configuration uses a 15-parameter hybrid equilibrium chemistry model that includes the species $\mathrm{H_2O}$, $\mathrm{CO_2}$, $\mathrm{CO}$, $\mathrm{SO_2}$, $\mathrm{Na}$, and $\mathrm{K}$, together with the aerosol species $\mathrm{MgSiO_3}$ and $\mathrm{ZnS}$, a uniform cloud prescription, and a Madhusudhan--Seager P--T profile. \tr{Unlike the standard equilibrium chemistry model, where
chemical abundances are determined by thermochemical equilibrium,
the hybrid equilibrium model offers greater flexibility by
allowing selected species, such as $\mathrm{SO_2}$, to deviate
from equilibrium, thereby accounting for disequilibrium processes.
Further details of the implementation are provided in
\citet{Deka_2025}.} \texttt{PandExo}-based noise realizations were incorporated into the synthetic spectra before training. The adopted parameter ranges are provided in Tables~\ref{tab:MIRI_synth_comparison} and~\ref{tab:MIRI_noise_comparison} for the MIRI/LRS experiments and in Table~\ref{tab:hybrid_comparison} for the full JWST retrieval.

We performed hyperparameter optimization for the \tr{TcNF} architecture. The explored ranges included 4--12 flow layers, 4--8 attention heads, 40--200 training epochs, learning rates between $10^{-6}$ and $10^{-4}$, and batch sizes between 64 and 256. The optimal configuration depends on the size of the training dataset and the dimensionality and complexity of the parameter space. For the two-million-spectrum training datasets, we trained the models for 100 epochs and selected the configuration that minimized the validation loss. The adopted architecture uses six flow layers, four attention heads, a batch size of 128, and a learning rate of $5\times10^{-4}$. The model employs the Gaussian Error Linear Unit (GELU) activation function \citep{Hendrycks2016GELU} within its multilayer perceptron (MLP) layers, the AdamW optimizer \citep{Loshchilov2019AdamW}, and a \texttt{ReduceLROnPlateau} scheduler for adaptive learning-rate control.

The objective of \texttt{NEXONN} is to learn the conditional posterior distribution $\mathrm{p(\theta\mid\mathbf{x}})$ by optimizing the parameters of the \tr{TcNF}. For the affine transformations described in Equations~\ref{eq:affine1} and~\ref{eq:affine2}, the \tr{transformer} provides the conditioning information used to determine the scale, $s$, and shift, $t$, functions of the flow. During inference, samples are drawn from the base distribution $\mathrm{p(\mathbf{z}})$ and transformed through the inverse flow, conditioned on the observed spectrum $\mathbf{x}$, to generate samples from the learned posterior $\mathrm{p(\theta\mid\mathbf{x}})$. These samples are then used to construct the marginalized and joint posterior distributions, which are visualized using corner plots. The widths and shapes of the resulting posterior distributions characterize the uncertainties inferred by the trained model under the adopted forward model, training distribution, and observational noise model.
 
\subsubsection{\textbf{Statistical Assessment of Retrieved Posteriors}}
\label{statistical}

To quantitatively assess the agreement between the posterior distributions obtained with \texttt{NEXONN} and \texttt{PyMultiNest}, we employ several complementary statistical measures, including the \tr{Kullback--Leibler (KL) divergence, Jensen--Shannon (JS) divergence and standardised Wasserstein distance (SWD)}, and posterior overlap (\%). These metrics are evaluated separately for the one-dimensional marginalized posterior distribution of each retrieved parameter. Together, they characterize different aspects of posterior agreement, including information divergence, overall distributional similarity, geometric displacement of probability mass, and shared probability mass.

For the KL divergence, we treat the \texttt{PyMultiNest}
posterior as the reference distribution, $\mathrm{P}$,
and the corresponding \texttt{NEXONN} posterior as
$\mathrm{Q}$. The KL divergence measures the information
loss incurred when $\mathrm{Q}$ is used to approximate
$\mathrm{P}$ and is defined as

\begin{equation}
    \mathrm{D}_{\mathrm{KL}}(\mathrm{P} \parallel \mathrm{Q})
    =
    \int
    p(x)
    \log\left[
    \frac{p(x)}{q(x)}
    \right]
    \mathrm{d}x,
\end{equation}

where $p(x)$ and $q(x)$ denote the corresponding
marginalized probability density functions and $x$
denotes the retrieved parameter under consideration.
\tr{To avoid numerical singularities when $p(x)>0$
and $q(x)=0$, we apply a small numerical threshold
of $\epsilon=10^{-15}$ to both probability densities
before evaluating the KL divergence.}
Since the KL divergence is generally asymmetric,

\begin{equation}
    \mathrm{D}_{\mathrm{KL}}(\mathrm{P} \parallel \mathrm{Q})
    \neq
    \mathrm{D}_{\mathrm{KL}}(\mathrm{Q} \parallel \mathrm{P}),
\end{equation}

\tr{we additionally compute the symmetric JS divergence,}

\begin{equation}
    \mathrm{D}_{\mathrm{JS}}(\mathrm{P} \parallel \mathrm{Q})
    =
    \frac{1}{2}
    \mathrm{D}_{\mathrm{KL}}(\mathrm{P} \parallel \mathrm{M})
    +
    \frac{1}{2}
    \mathrm{D}_{\mathrm{KL}}(\mathrm{Q} \parallel \mathrm{M}),
\end{equation}

where

\begin{equation}
    \mathrm{M} = \frac{1}{2}(\mathrm{P}+\mathrm{Q})
\end{equation}

is the mixture distribution of $\mathrm{P}$ and
$\mathrm{Q}$. \tr{The JS divergence remains finite
even when one distribution vanishes where the other
is nonzero, since the mixture distribution is positive
wherever either distribution is positive.} 

The Wasserstein distance (WD) provides a complementary
geometric measure of the difference between two distributions
by quantifying the minimum transport cost required to transform
one probability distribution into the other. The \tr{WD} of order $p$ is defined as

\begin{equation}
    \mathrm{W}_{p}(\mathrm{P},\mathrm{Q})
    =
    \left[
    \inf_{\gamma \in \Gamma(\mathrm{P},\mathrm{Q})}
    \int
    \|x-y\|^{p}
    \,\mathrm{d}\gamma(x,y)
    \right]^{1/p},
\end{equation}

where $\Gamma(\mathrm{P},\mathrm{Q})$ denotes the set of
all joint distributions with marginals $\mathrm{P}$ and
$\mathrm{Q}$, and $x$ and $y$ denote samples from the
respective marginalized posterior distributions.
\tr{Since WD depends on the scale and units of the retrieved parameters, direct comparisons across different parameters may be misleading. To address this limitation, we use the standardized WD. For each parameter, we calculate the mean and standard deviation of the combined posterior samples and then perform z-score normalization before calculating WD. This provides a dimensionless measure that enables more meaningful comparisons across different parameters.}

The posterior overlap provides an additional measure of similarity by quantifying the common probability mass shared by the two marginalized posterior distributions. We define the fractional posterior overlap as

\begin{equation}
    \mathcal{O}(\mathrm{P},\mathrm{Q})
    =
    \int
    \min\left[p(x),q(x)\right]
    \mathrm{d}x,
\end{equation}

with $\mathcal{O}=1$ corresponding to complete overlap and $\mathcal{O}=0$ corresponding to no overlap. The values reported in this work are expressed as percentages, $100\,\mathcal{O}$.

These metrics emphasize different properties of posterior agreement and should therefore be interpreted jointly. The KL divergence satisfies $\mathrm{D}_{\mathrm{KL}}(\mathrm{P}\parallel\mathrm{Q})\geq0$, with $\mathrm{D}_{\mathrm{KL}}=0$ only when the two distributions are identical almost everywhere. Smaller values indicate greater similarity relative to the chosen reference distribution, while larger values indicate greater differences in the probability density assigned by the two posteriors. The JS divergence provides a symmetric measure of overall distributional difference, with smaller values indicating greater similarity between the distributions. \tr{The standardized Wasserstein distance (SWD) measures the geometric displacement of probability mass relative to the combined posterior spread and is sensitive to differences in the locations and shapes of the distributions, with smaller SWD values indicating greater posterior similarity.}  Finally, larger posterior-overlap values indicate a greater fraction of shared probability mass. Consequently, similar median parameter estimates do not necessarily imply similar posterior distributions, and the combined use of these metrics provides a more complete assessment of posterior agreement than comparison of central values alone. The statistical quantities reported in this work were calculated from the marginalized posterior distributions using the relevant routines available in \texttt{SciPy}.

In addition to these posterior-comparison metrics, we
independently assess whether the credible regions inferred
by \texttt{NEXONN} are statistically calibrated using
coverage tests. Following
\citet{lemos2023samplingL,Vasist_2023},
\tr{we evaluate the empirical coverage probability using
repeated realizations of simulated spectra with known
ground truths. For each realization, we construct the
marginalized posterior distribution of each parameter
and check whether its true value lies within the
corresponding credible interval at a nominal credible
level of $1-\alpha$, where $\alpha$ represents the
fraction of posterior probability outside the interval.
For a well-calibrated posterior, a credible interval
containing a fraction $1-\alpha$ of the posterior
probability should contain the true parameter value
in the same fraction of realizations. Thus, the
empirical coverage probability should follow the
ideal one-to-one relation,}

\begin{equation}
    C_{\mathrm{empirical}} = C_{\mathrm{nominal}},
\end{equation}

corresponding to a slope of $m=1$. Empirical coverage below the nominal credible level indicates an overconfident posterior, for which the credible regions are too narrow, whereas coverage above the nominal credible level indicates a conservative posterior, for which the credible regions are broader than required for exact calibration. We additionally calculate the coefficient of determination, $R^2$, to quantify how closely the empirical coverage follows the fitted calibration relation. Values of $R^2$ close to unity indicate that the empirical coverage is well described by the fitted relation, while the slope $m$ provides the primary diagnostic of whether the inferred credible regions are calibrated, overconfident, or conservative.

\begin{figure*}
    \centering
    \includegraphics[width=\linewidth]{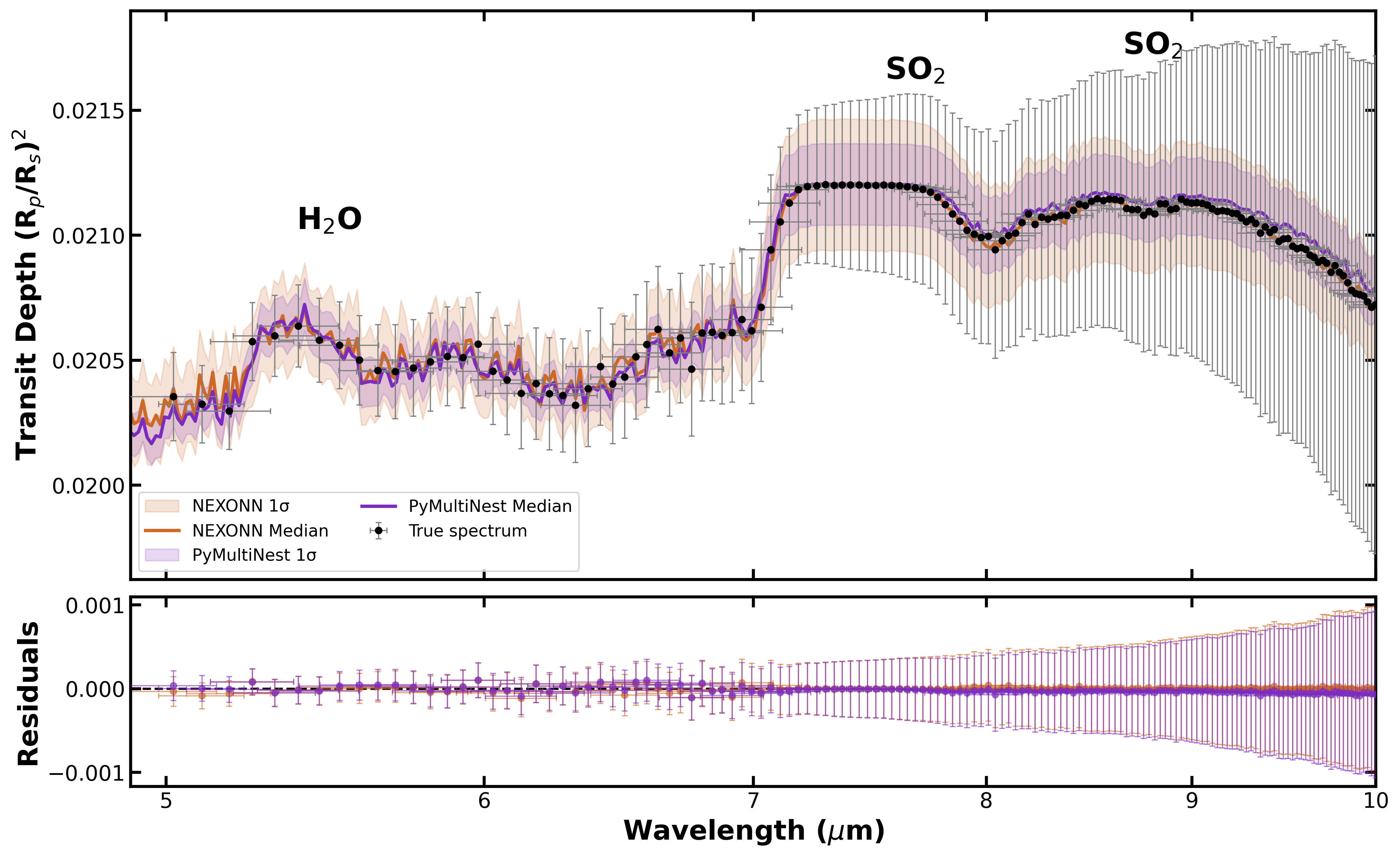}
    \caption{Comparison of the transmission spectra corresponding to the \texttt{PyMultiNest} (purple) and \texttt{NEXONN} (orange) retrievals for the simulated JWST MIRI/LRS experiment. The simulated observations were generated using the \texttt{NEXOTRANS} forward model with \texttt{PandExo}-based noise. Both retrievals are performed on the same simulated dataset with known input atmospheric parameters.}
    \label{fig:simulated_spectrum}
\end{figure*}

\section{\textbf{Validation and Results}}
\label{sec:results}

In this section, we present the retrieval results obtained using the \texttt{NEXONN} framework. We first validate \texttt{NEXONN} using a simulated spectrum with known atmospheric parameters. We then apply the framework to the JWST MIRI/LRS transmission spectrum of WASP-39~b spanning $5$--$12\,\mu\mathrm{m}$, followed by a retrieval using the combined NIRISS/SOSS, NIRSpec/PRISM, and MIRI/LRS observations spanning approximately $0.6$--$12\,\mu\mathrm{m}$. In each case, the resulting \texttt{NEXONN} posterior distributions are compared against those obtained with \texttt{PyMultiNest} within the \texttt{NEXOTRANS} framework.

\begin{figure*}
    \centering
    \includegraphics[width=\linewidth]{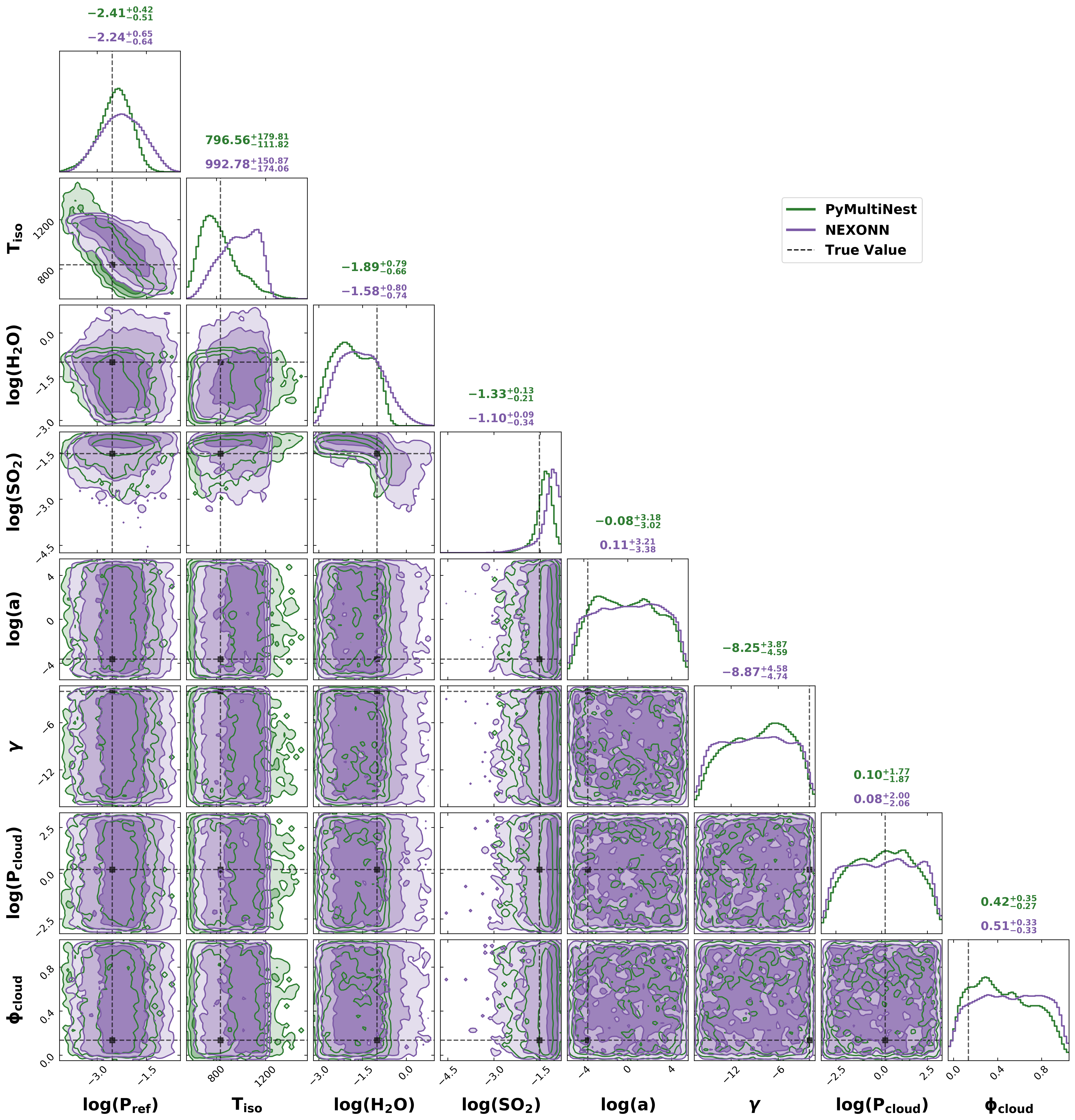}
    \caption{Comparison of the posterior distributions retrieved with \texttt{PyMultiNest} (green) and \texttt{NEXONN} (purple) for the simulated spectrum generated using the \texttt{NEXOTRANS} forward model with \texttt{PandExo}-based noise. The black markers indicate the true values of the atmospheric parameters used to generate the simulated spectrum.}
    \label{synthetic}
\end{figure*}

\subsection{\textbf{Benchmarking \texttt{NEXONN} against
\texttt{PyMultiNest} Using a Simulated Spectrum with
Known Atmospheric Parameters}}
\label{sim-retrieval}

We first validate \texttt{NEXONN} by retrieving the known atmospheric parameters of a simulated spectrum. The spectrum was generated using the \texttt{NEXOTRANS} forward model with the atmospheric configuration described in Section~\ref{Training} over the JWST MIRI/LRS wavelength range. \texttt{PandExo}-based noise was subsequently incorporated using the observational configuration adopted for WASP-39~b. The resulting simulated spectrum was then retrieved independently using both \texttt{NEXONN} and \texttt{PyMultiNest}.

Figure~\ref{fig:simulated_spectrum} compares the transmission spectra corresponding to the \texttt{NEXONN} and \texttt{PyMultiNest} retrievals with the simulated observations. Both retrievals reproduce the simulated spectrum across the MIRI/LRS wavelength range and recover the principal spectral features within the observational uncertainties. The \texttt{NEXONN} and \texttt{PyMultiNest} median spectra closely follow one another, and their uncertainty envelopes show substantial agreement across the wavelength range. \tr{This indicates that both retrieval methods provide comparably good fits to the simulated observations in spectral space.} Spectral agreement alone, however, does not establish agreement of the underlying atmospheric posteriors because different combinations of atmospheric parameters can produce similar transmission spectra. We therefore assess the marginalized and joint posterior distributions in Figure~\ref{synthetic} and quantify their agreement using the statistical measures reported in Table~\ref{tab:MIRI_synth_comparison}.

\tr{Figure~\ref{synthetic} presents the posterior distributions, while Table~\ref{tab:MIRI_synth_comparison} summarizes the known input values, retrieved constraints, and statistical agreement between the one-dimensional marginalized \texttt{NEXONN} and \texttt{PyMultiNest} posteriors
for all eight atmospheric parameters.} The comparison provides two complementary tests. First, because the generating parameters are known, it tests whether the retrievals remain consistent with the underlying atmospheric state. Second, the KL divergence, JS divergence, \tr{SWD}, and posterior overlap quantify how closely \texttt{NEXONN} reproduces the corresponding \texttt{PyMultiNest} marginalized
posterior distributions. Importantly, close agreement between the two retrieval methods does not necessarily imply that an individual parameter is tightly constrained by the spectrum.

\tr{For the logarithmic reference pressure, $\log(\mathrm{P_{\mathrm{ref}}})$, the two methods yield consistent posterior constraints, with $\mathrm{D}_{\mathrm{KL}}=0.16$, $\mathrm{D}_{\mathrm{JS}}=0.07$, an SWD of $0.37$, and a posterior overlap of $80.32\%$. The known input value, $\log(\mathrm{P_{\mathrm{ref}}})=-2.55$, lies within the $68\%$ credible intervals of both retrievals. In contrast, the isothermal temperature, $\mathrm{\mathrm{T_{\mathrm{iso}}}}$, shows greater differences between the marginalized posteriors, with an overlap of $54.69\%$, $\mathrm{D}_{\mathrm{KL}}=0.81$, $\mathrm{D}_{\mathrm{JS}}=0.21$, and an SWD of $0.96$. The posterior medians differ appreciably, with \texttt{NEXONN} yielding $992.78\,\mathrm{K}$ compared with $796.56\,\mathrm{K}$ from \texttt{PyMultiNest}. Nevertheless, the known input temperature, $\mathrm{T_{\mathrm{iso}}}=832.68\,\mathrm{K}$, lies within the $68\%$ credible intervals inferred by both methods. This illustrates that recovering the known input value within the inferred uncertainties does not necessarily imply close agreement between the posterior distributions.}

\tr{For $\log(\mathrm{H_2O})$, \texttt{NEXONN} yields $-1.58^{+0.80}_{-0.74}$ compared with $-1.89^{+0.79}_{-0.66}$ from \texttt{PyMultiNest}, with $\mathrm{D}_{\mathrm{KL}}=0.15$, $\mathrm{D}_{\mathrm{JS}}=0.07$, an SWD of $0.44$, and a posterior overlap of $85.06\%$. This indicates reasonably good agreement between the marginalized water-abundance posteriors. In contrast, $\log(\mathrm{SO_2})$ shows lower posterior agreement, with $\mathrm{D}_{\mathrm{KL}}=0.70$, $\mathrm{D}_{\mathrm{JS}}=0.26$, an SWD of $0.56$, and a posterior overlap of $50.16\%$. The known input abundance, $\log(\mathrm{SO_2})=-1.54$, lies outside the reported $68\%$ credible interval of the \texttt{NEXONN} retrieval, which yields $-1.10^{+0.09}_{-0.34}$, while the \texttt{PyMultiNest} estimate of $-1.33^{+0.13}_{-0.21}$ places the generating value at the lower boundary of its reported $68\%$ interval. The \texttt{NEXONN} posterior is also concentrated closer to the upper limit of the adopted abundance range, suggesting that posterior approximation near parameter boundaries warrants further investigation. These results indicate that the agreement
between the TcNF approximation and the \texttt{PyMultiNest} posterior is parameter dependent, even for the controlled simulated-spectrum experiment. The known input abundances provide an independent reference for assessing the retrieval constraints, although agreement with the generating values should be distinguished from agreement between the inferred posterior distributions.}

A different behavior is seen for the cloud and haze parameters, and this provides an important test of the probabilistic nature of \texttt{NEXONN}. The marginalized posteriors for $\log(a)$, $\gamma$, $\log(\mathrm{P_{cloud}})$, and $\phi_{\mathrm{cloud}}$ are broad, indicating that these parameters are only weakly constrained by the simulated MIRI/LRS spectrum. Nevertheless, \texttt{NEXONN} closely reproduces the corresponding \texttt{PyMultiNest} posteriors. \tr{The posterior overlaps are $94.92\%$, $93.61\%$, $93.46\%$, and $90.32\%$ for $\log(a)$, $\gamma$, $\log(\mathrm{P_{cloud}})$, and $\phi_{\mathrm{cloud}}$, respectively. Their KL divergences are $0.01$, $0.01$, $0.01$, and $0.03$, while the corresponding JS divergences are $0.00$, $0.01$, $0.00$, and $0.01$ at the precision reported in Table~\ref{tab:MIRI_synth_comparison}.
The corresponding SWD values are $0.07$, $0.12$, $0.07$, and $0.21$, respectively. The consistently high posterior overlaps and small divergences indicate that \texttt{NEXONN} closely reproduces the broad \texttt{PyMultiNest} posteriors for these weakly constrained parameters. However, high posterior overlap should not be interpreted as precise recovery of the cloud and haze properties. Instead, it demonstrates that the TcNF can reproduce the limited information content of the simulated spectrum by retaining broad posterior uncertainties.}

Figure~\ref{synthetic} provides an additional view of this behavior. The marginalized distributions show the \tr{parameter-dependent agreement} quantified in Table~\ref{tab:MIRI_synth_comparison}, while the \tr{two-dimensional projections provide a qualitative comparison of the inferred parameter correlations and degeneracies, although the degree of agreement varies among parameter pairs.} Because the statistical quantities in
Table~\ref{tab:MIRI_synth_comparison} are calculated from one-dimensional marginalized posteriors, the corner plot provides a complementary visual assessment of whether \texttt{NEXONN} also preserves the principal correlations present in the \texttt{PyMultiNest} solution. \tr{The qualitative correspondence between some of the two-dimensional contours suggests that the TcNF captures aspects of the parameter correlations present in the \texttt{PyMultiNest} solution. However, the one-dimensional comparison metrics do not provide a quantitative validation of the full multidimensional posterior structure.}

\tr{Taken together, the simulated-spectrum experiment provides a controlled benchmark at three complementary levels. First, Figure~\ref{fig:simulated_spectrum} shows that \texttt{NEXONN} and \texttt{PyMultiNest} recover closely matching spectral solutions.
Second, the known input parameters provide an independent reference for assessing the retrieved
atmospheric constraints. Third, the KL divergence, JS divergence, SWD, and posterior-overlap statistics reveal parameter-dependent agreement between the marginalized posterior distributions. While the cloud and haze parameters show consistently high posterior overlaps, appreciable differences are found for $\mathrm{\mathrm{T_{\mathrm{iso}}}}$ and $\log(\mathrm{SO_2})$. These results show that \texttt{NEXONN} can reproduce broad posterior distributions for weakly constrained cloud and haze parameters while recovering spectral solutions consistent with the simulated observations. However, the appreciable posterior discrepancies for $\mathrm{\mathrm{T_{\mathrm{iso}}}}$ and $\log(\mathrm{SO_2})$ demonstrate that accurate spectral reconstruction does not guarantee accurate posterior estimation for every atmospheric parameter. The experiment therefore provides a useful in-distribution benchmark of the TcNF framework while identifying specific limitations that motivate further improvements in posterior
approximation and multidimensional validation.}

\begin{table*}[ht]
\centering
\caption{\tr{Comparative overview on the retrieved parameters using \texttt{PyMultiNest} and \texttt{NEXONN} for simulated spectrum. The model considers a patchy-gray cloud, isothermal P-T profile and Free chemistry.}}
\label{tab:MIRI_synth_comparison}

\small 
\renewcommand{\arraystretch}{1.3}
\setlength{\tabcolsep}{2pt} 

\begin{tabular}{@{} l c c c c @{\hspace{0.1pt}} c @{\hspace{0.1pt}} c @{\hspace{0.1pt}} c @{\hspace{0.1pt}} c @{}}
\noalign{\hrule height 0.8pt \vskip 1.2pt \hrule height 0.8pt}

Parameter & 
\begin{tabular}{@{}c@{}}True \\[-0.5ex] Value\end{tabular} & 
\begin{tabular}{@{}c@{}}Training \\[-0.5ex] Range$^{*}$ \end{tabular} & 
\texttt{PyMultiNest} & 
\texttt{NEXONN} & 
\begin{tabular}{@{}c@{}}KL \\[-0.5ex] Divergence\end{tabular} & 
\begin{tabular}{@{}c@{}}JS \\[-0.5ex] Divergence\end{tabular} & 
\begin{tabular}{@{}c@{}}Std. Wasserstein \\[-0.5ex] Distance \end{tabular} & 
\begin{tabular}{@{}c@{}}Overlap \\[-0.5ex] (\%)\end{tabular} \\
\hline

$\log(\mathrm{P_{\mathrm{ref}}})$ 
& -2.55
& $\mathcal{U}(-4.00,-0.50)$
& $-2.41^{+0.42}_{-0.51}$ 
& $-2.24^{+0.65}_{-0.64}$ 
& 0.16 & 0.07 & 0.37 & 80.32 \\

$\mathrm{\mathrm{T_{\mathrm{iso}}}}$ 
& 832.68
& $\mathcal{U}(600.00,1500.00)$
& $796.56^{+179.81}_{-111.82}$ 
& $992.78^{+150.87}_{-174.06}$ 
& 0.81 & 0.21 & 0.96 & 54.69 \\

$\log(\mathrm{H_2O})$ 
& -1.00
& $\mathcal{U}(-3.00,1.00)$
& $-1.89^{+0.79}_{-0.66}$ 
& $-1.58^{+0.80}_{-0.74}$ 
& 0.15 & 0.07 & 0.44 & 85.06 \\

$\log(\mathrm{SO_2})$ 
& -1.54
& $\mathcal{U}(-7.00,-1.00)$
& $-1.33^{+0.13}_{-0.21}$ 
& $-1.10^{+0.09}_{-0.34}$ 
& 0.70 & 0.26 & 0.56 & 50.16 \\

$\log(a)$ 
& -3.63
& $\mathcal{U}(-5.00,5.00)$
& $-0.08^{+3.18}_{-3.02}$ 
& $0.11^{+3.21}_{-3.38}$ 
& 0.01 & 0.00 & 0.07 & 94.92 \\

$\gamma$ 
& -2.03
& $\mathcal{U}(-16.00,-2.00)$
& $-8.25^{+3.87}_{-4.59}$ 
& $-8.87^{+4.58}_{-4.74}$ 
& 0.01 & 0.01 & 0.12 & 93.61 \\

$\log(\mathrm{P_{\mathrm{cloud}}})$ 
& 0.20
& $\mathcal{U}(-3.00,3.00)$
& $0.10^{+1.77}_{-1.87}$ 
& $0.08^{+2.00}_{-2.06}$ 
& 0.01 & 0.00 & 0.07 & 93.46 \\

$\mathrm{\phi_{\mathrm{cloud}}}$ 
& 0.14
& $\mathcal{U}(0.00,1.00)$
& $0.42^{+0.35}_{-0.27}$ 
& $0.51^{+0.33}_{-0.33}$ 
& 0.03 & 0.01 & 0.21 & 90.32 \\

\hline\hline
\end{tabular}

\vspace{1mm}
\begin{flushleft}
{\footnotesize $^{*}$ The training range represents the parameter ranges over which the \texttt{NEXONN} model was trained and the same range was used as the prior for the \texttt{PyMultiNest} retrieval. $\mathrm{log(a), \gamma, log(P_{cloud}), \phi_{cloud}}$ are the cloud parameters for patchy-deck cloud and haze considerations.}
\end{flushleft}

\end{table*}

\begin{figure*}
    \centering
    \includegraphics[width=\linewidth]{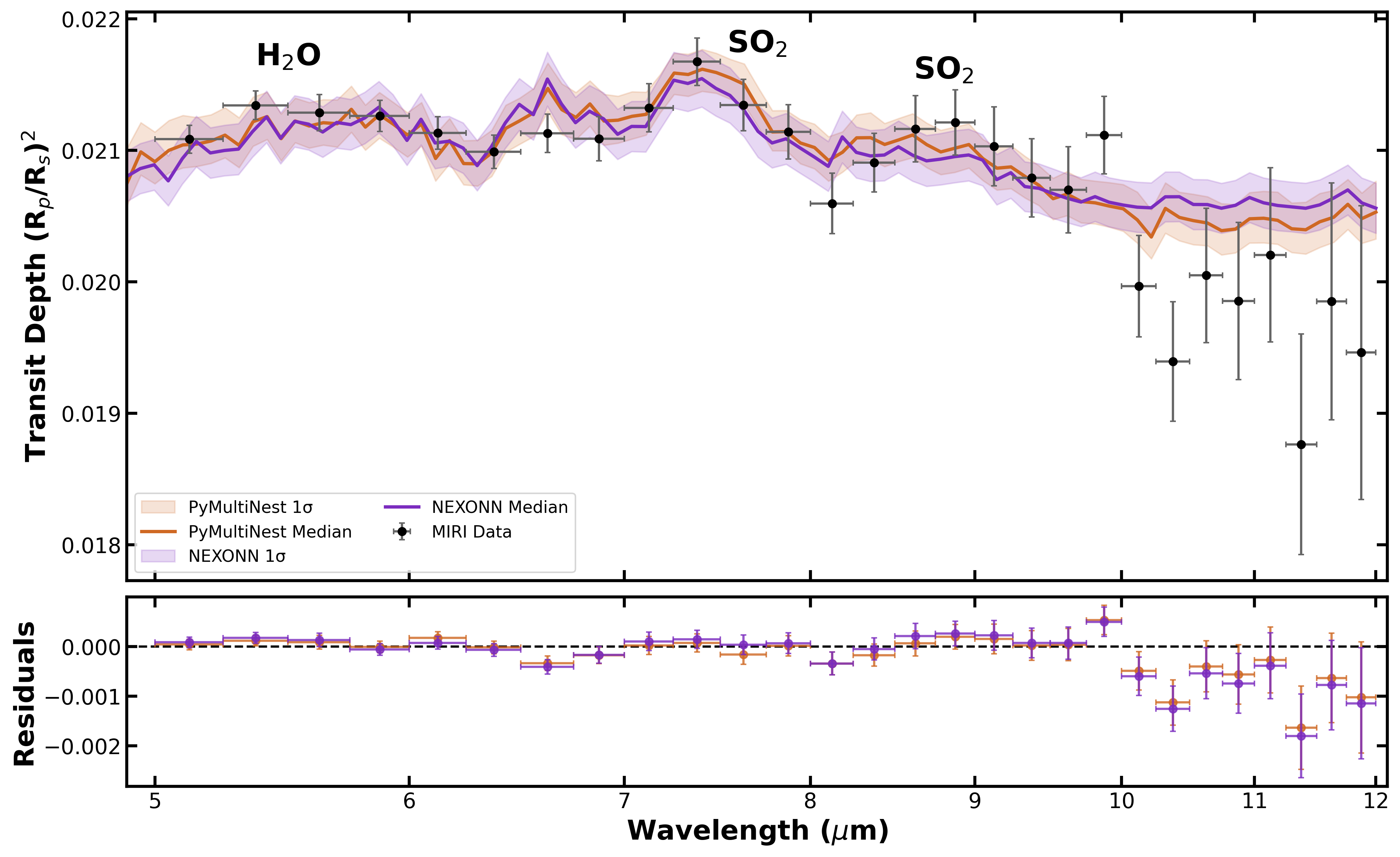}
    \caption{Comparison of the retrieved JWST MIRI/LRS transmission spectra of WASP-39~b using \texttt{PyMultiNest} (orange) and \texttt{NEXONN} (purple). The observations are based on the \citet{Powell2024} reduction performed with the \texttt{Eureka!} pipeline. The retrieval model assumes Free Chemistry, an isothermal pressure--temperature (P--T) profile, haze, and patchy-gray clouds. The 28 observed spectral points are shown in black. The lower panel shows the residuals relative to the observed transit depths.}
    \label{fig:miri spectrum}
\end{figure*}

\begin{figure*}
    \centering
    \includegraphics[width=1\linewidth]{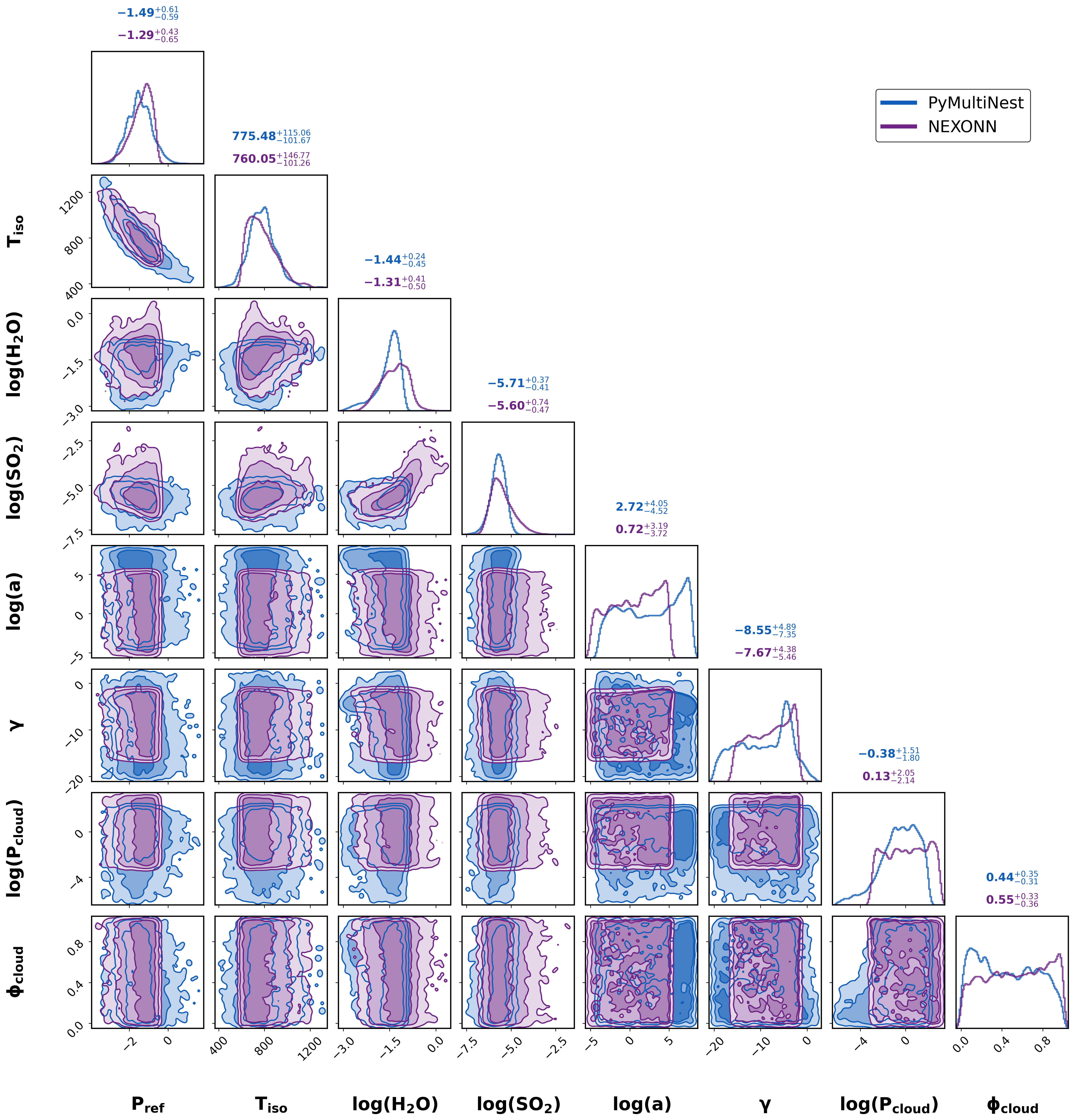}
    \caption{The figure shows the corner plots comparing the posterior distributions retrieved using \texttt{PyMultiNest} (blue), \texttt{NEXONN} with \texttt{PandExo} noise (purple) for JWST MIRI (5 - 12$\mu$m) observed spectrum of WASP-39 b, reduced using \texttt{Eureka!} pipeline \citep{Powell2024}.}
    \label{fig:miri_corner}
\end{figure*}

\begin{table*}[ht]
\centering
\caption{\tr{Comparative overview on the retrieved parameters using \texttt{PyMultiNest} and \texttt{NEXONN} for WASP-39 b, JWST MIRI (5--12$\mu$m range). The model considers a patchy-gray cloud, isothermal P-T profile and Free chemistry.}}
\label{tab:MIRI_noise_comparison}

\small 
\renewcommand{\arraystretch}{1.3}
\setlength{\tabcolsep}{4.5pt} 

\begin{tabular}{@{} l c c c c c c c @{}}
\noalign{\hrule height 0.8pt \vskip 1.2pt \hrule height 0.8pt}

Parameter & 
\begin{tabular}{@{}c@{}}Training \\[-0.5ex] Range$^{*}$ \end{tabular} & 
\texttt{PyMultiNest} & 
\texttt{NEXONN} & 
\begin{tabular}{@{}c@{}}KL \\[-0.5ex] Divergence\end{tabular} & 
\begin{tabular}{@{}c@{}}JS \\[-0.5ex] Divergence\end{tabular} & 
\begin{tabular}{@{}c@{}}Std. Wasserstein \\[-0.5ex] Distance \end{tabular} & 
\begin{tabular}{@{}c@{}}Overlap \\[-0.5ex] (\%)\end{tabular} \\
\hline

$\log(\mathrm{P_{\mathrm{ref}}})$ 
& $\mathcal{U}(-4.0,-0.5)$
& $-1.49^{+0.61}_{-0.59}$ 
& $-1.29^{+0.43}_{-0.65}$
& 0.13 & 0.05 & 0.25 & 81.11 \\

$\mathrm{\mathrm{T_{\mathrm{iso}}}}$ 
& $\mathcal{U}(600,1500)$
& $775.48^{+115.06}_{-101.67}$ 
& $760.05^{+146.77}_{-101.26}$
& 0.08 & 0.03 & 0.15 & 87.03 \\

$\log(\mathrm{H_2O})$ 
& $\mathcal{U}(-3,1)$
& $-1.44^{+0.24}_{-0.45}$ 
& $-1.31^{+0.41}_{-0.50}$
& 1.67 & 0.14 & 0.48 & 71.97 \\

$\log(\mathrm{SO_2})$ 
& $\mathcal{U}(-7,-1)$
& $-5.71^{+0.37}_{-0.41}$ 
& $-5.60^{+0.74}_{-0.47}$
& 1.52 & 0.09 & 0.47 & 79.41 \\

$\log(a)$ 
& $\mathcal{U}(-5,5)$
& $2.72^{+4.05}_{-4.52}$ 
& $0.72^{+3.19}_{-3.72}$
& 0.41 & 0.17 & 0.61 & 70.68 \\

$\gamma$ 
& $\mathcal{U}(-16,-2)$
& $-8.55^{+4.89}_{-5.35}$ 
& $-7.67^{+4.38}_{-5.46}$
& 0.19 & 0.09 & 0.27 & 79.87 \\

$\log(\mathrm{P_{\mathrm{cloud}}})$ 
& $\mathcal{U}(-3,3)$
& $-0.38^{+1.51}_{-1.80}$ 
& $0.13^{+2.05}_{-2.14}$
& 0.67 & 0.11 & 0.37 & 77.73 \\

$\phi_{\mathrm{cloud}}$ 
& $\mathcal{U}(0,1)$
& $0.44^{+0.35}_{-0.31}$ 
& $0.55^{+0.33}_{-0.36}$
& 0.06 & 0.02 & 0.30 & 89.02 \\

\hline\hline
\end{tabular}

\vspace{1mm}
\begin{flushleft}
{\footnotesize $^{*}$ The training range represents the parameter ranges over which the \texttt{NEXONN} model was trained and the same range was used as the prior for the \texttt{PyMultiNest} retrieval. $\mathrm{log(a), \gamma, log(P_{cloud}), \phi_{cloud}}$ are the cloud parameters for patchy-deck cloud and haze considerations. }
\end{flushleft}

\end{table*}

\begin{figure*}
    \centering
    \includegraphics[width=\linewidth]{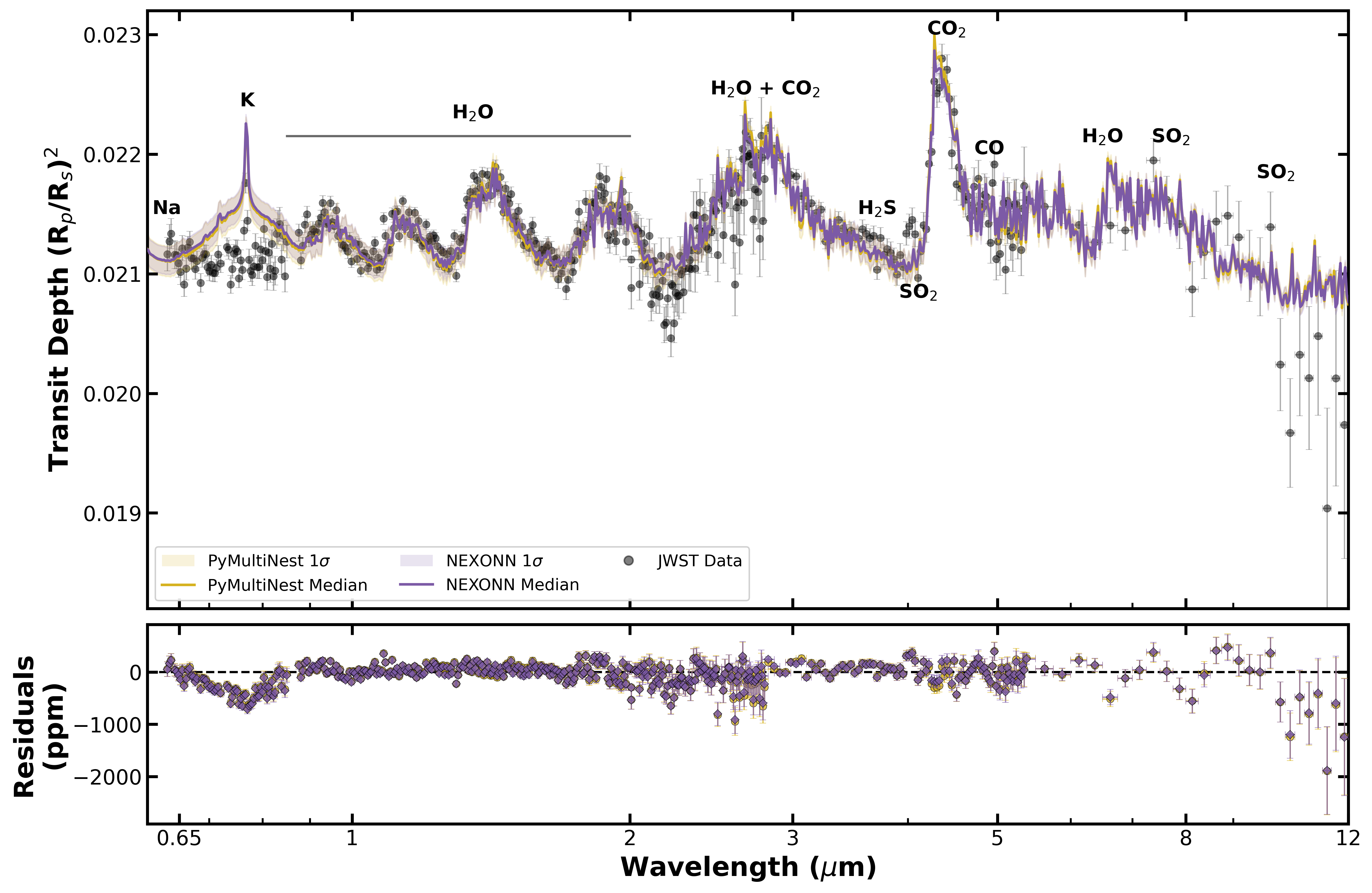}
    \caption{Comparison of the retrieved transmission spectra of WASP-39~b over the full JWST wavelength range of $0.6$--$12\,\mu\mathrm{m}$ using \texttt{PyMultiNest} (yellow) and \texttt{NEXONN} (purple). The \texttt{NEXONN} model was trained using \texttt{PandExo}-based noise realizations. The retrieval adopts the Hybrid Equilibrium Chemistry model together with a uniform cloud and aerosol prescription.}
    \label{fig:hybrid spectrum}
\end{figure*}

\begin{figure*}
    \centering
    \includegraphics[width=\linewidth]{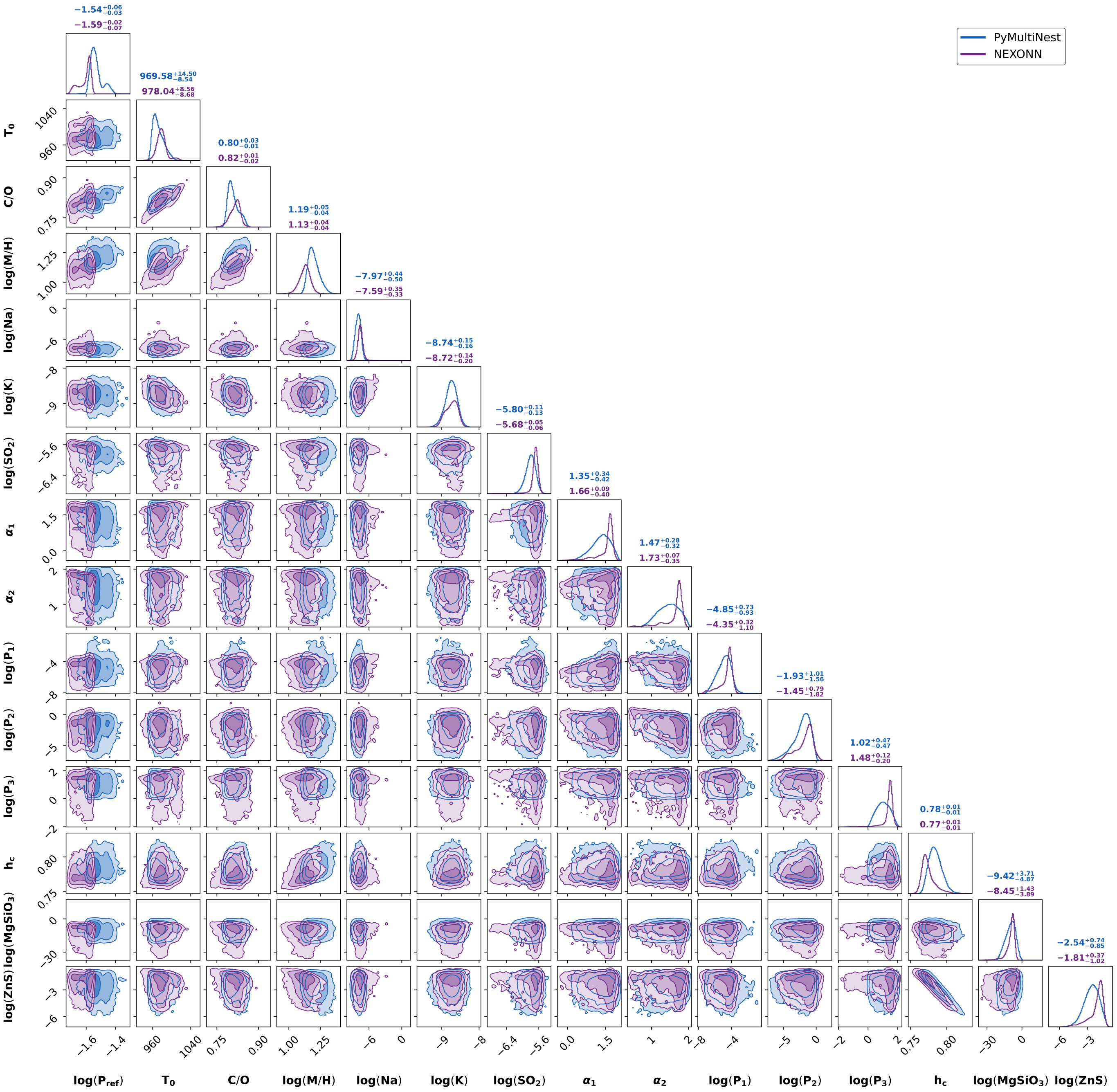}
    \caption{Comparison of the posterior distributions obtained with \texttt{PyMultiNest} (blue) and \texttt{NEXONN} (purple) for the full JWST transmission spectrum of WASP-39~b spanning $0.6$--$12\,\mu\mathrm{m}$.}
    \label{fig:hybrid corner plot}
\end{figure*}

\subsection{\textbf{Applying \texttt{NEXONN} to the JWST MIRI/LRS Retrieval of WASP-39~b}}
\label{miri-retrieval}

We now apply \texttt{NEXONN} to the JWST MIRI/LRS transmission spectrum of WASP-39~b spanning $5$--$12\,\mu\mathrm{m}$. The spectrum was obtained from the NASA Exoplanet Archive atmospheric spectroscopy table\footnote{\url{https://exoplanetarchive.ipac.caltech.edu/cgi-bin/atmospheres/nph-firefly?atmospheres}} and corresponds to the \citet{Powell2024} reduction performed using the \texttt{Eureka!} pipeline. We benchmark the \texttt{NEXONN} retrieval against the \texttt{PyMultiNest} retrieval presented by \citet{Deka_2025}, adopting the same best-fit atmospheric model for the MIRI/LRS wavelength range, which yielded a reduced $\chi^2$ of 1.15. The model assumes Free Chemistry, an isothermal pressure--temperature (P--T) profile, haze, and patchy-gray clouds. The isothermal temperature, logarithmic abundances of $\mathrm{H_2O}$ and $\mathrm{SO_2}$, and the cloud and haze parameters are treated as free parameters.

Unlike the simulated-spectrum experiment in Section~\ref{sim-retrieval}, the atmospheric parameters underlying the observed WASP-39~b spectrum are not known. The present experiment therefore tests whether \texttt{NEXONN} can reproduce the posterior constraints obtained with the conventional \texttt{PyMultiNest} retrieval when applied to real JWST observations. \tr{Because both methods employ the same atmospheric forward model and parameter ranges, this comparison primarily assesses the agreement between their inferred posterior distributions under the adopted modelling assumptions, rather than establishing the accuracy of the atmospheric model itself.} The statistical quantities in Table~\ref{tab:MIRI_noise_comparison} compare the one-dimensional marginalized posterior distributions parameter by parameter, while Figure~\ref{fig:miri_corner} provides a complementary view of the joint posterior structure.

The reference pressure and atmospheric temperature show relatively strong agreement between the two retrieval methods. For $\tr{\log(\mathrm{P_{\mathrm{ref}}})}$, \texttt{NEXONN} yields $-1.29^{+0.43}_{-0.65}$ compared with $-1.49^{+0.61}_{-0.59}$ from \texttt{PyMultiNest}. The corresponding statistical measures are $\mathrm{D}_{\mathrm{KL}}=0.13$, $\mathrm{D}_{\mathrm{JS}}=0.05$, a \tr{SWD of $0.25$}, and a posterior overlap of $81.11\%$. For $\mathrm{T_{iso}}$, the inferred values are $760.05^{+146.77}_{-101.26}\,\mathrm{K}$ and $775.48^{+115.06}_{-101.67}\,\mathrm{K}$ for \texttt{NEXONN} and \texttt{PyMultiNest}, respectively. The posterior overlap increases to $87.03\%$, with $\mathrm{D}_{\mathrm{KL}}=0.08$, $\mathrm{D}_{\mathrm{JS}}=0.03$ and \tr{an SWD of $0.15$.} Together, these statistics indicate that the two methods infer similar reference-pressure and temperature constraints from the MIRI/LRS spectrum. \tr{The relatively small differences in their posterior medians, together with the high overlaps and low divergence measures, indicate that these parameters are among those for which the TcNF most closely approximates the corresponding \texttt{PyMultiNest} marginalized posteriors in this experiment.}

The molecular abundances provide a more instructive example of the distinction between agreement in central constraints and agreement in the detailed posterior density. \texttt{NEXONN} retrieves $\log(\mathrm{H_2O})=-1.31^{+0.41}_{-0.50}$ compared with $-1.44^{+0.24}_{-0.45}$ from \texttt{PyMultiNest}. Although these abundance constraints overlap substantially, the corresponding statistics are $\mathrm{D}_{\mathrm{KL}}=1.67$, $\mathrm{D}_{\mathrm{JS}}=0.14$, a \tr{SWD of $0.48$}, and a posterior overlap of $71.97\%$. Similarly, for $\log(\mathrm{SO_2})$, \texttt{NEXONN} yields $-5.60^{+0.74}_{-0.47}$ compared with $-5.71^{+0.37}_{-0.41}$ from \texttt{PyMultiNest}, with $\mathrm{D}_{\mathrm{KL}}=1.52$, $\mathrm{D}_{\mathrm{JS}}=0.09$, \tr{an SWD of $0.47$}, and a posterior overlap of $79.41\%$. \tr{The relatively small differences in the median abundances, $0.13$ dex for $\mathrm{H_2O}$ and $0.11$ dex for $\mathrm{SO_2}$, indicate that the two retrievals yield broadly consistent central abundance estimates. However, the larger KL divergences demonstrate that agreement in the central estimates does not extend uniformly to the full marginalized posterior densities.} \tr{Because KL divergence is asymmetric and sensitive to regions where the reference posterior assigns probability density that is relatively underrepresented by the approximating distribution, these larger values may arise from differences in posterior tails, widths, or detailed shapes. The smaller JS divergences and substantial posterior overlaps indicate that the distributions nevertheless share appreciable probability mass. These results therefore reveal parameter-dependent limitations in the TcNF posterior approximation rather than establishing that either retrieval provides a uniquely correct atmospheric solution.}

The cloud and haze parameters are generally more weakly constrained and show varying degrees of posterior agreement. The haze-amplitude parameter, $\log(a)$, has the lowest posterior overlap among the eight retrieved parameters, $70.68\%$, with $\mathrm{D}_{\mathrm{KL}}=0.41$, $\mathrm{D}_{\mathrm{JS}}=0.17$, and \tr{an SWD of $0.61$}. The corresponding marginalized posteriors are broad, with \texttt{PyMultiNest} yielding $2.72^{+4.05}_{-4.52}$ and \texttt{NEXONN} yielding $0.72^{+3.19}_{-3.72}$. \tr{The difference of $2.00$ dex between the posterior medians is appreciable, and Figure~\ref{fig:miri_corner} indicates that \texttt{NEXONN} assigns comparatively less probability density to the higher haze-amplitude region represented in the \texttt{PyMultiNest} posterior. Thus, despite the broad uncertainties, the agreement for $\log(a)$ is limited, particularly in the relative weighting of different regions of parameter space.} The haze-slope parameter, $\gamma$, is similarly broad, but shows somewhat greater posterior agreement, with an overlap of $79.87\%$, $\mathrm{D}_{\mathrm{KL}}=0.19$, $\mathrm{D}_{\mathrm{JS}}=0.09$, and \tr{an SWD of $0.27$}. These broad posterior distributions indicate limited constraining power on the detailed haze properties within the MIRI/LRS wavelength range. \tr{Nevertheless, the differences in posterior location and shape demonstrate that reproducing broad uncertainties alone is insufficient to establish accurate posterior approximation. The comparison therefore tests whether \texttt{NEXONN} reproduces both the extent and the probability-density structure of the weakly constrained haze posteriors, rather than whether these parameters are precisely determined by the observations.}

For the cloud-top pressure, $\tr{\log(\mathrm{P_{\mathrm{cloud}}})}$, the two retrievals yield broad distributions with a posterior overlap of $77.73\%$, together with $\mathrm{D}_{\mathrm{KL}}=0.67$, $\mathrm{D}_{\mathrm{JS}}=0.11$, and \tr{an SWD of $0.37$.} \tr{The posterior medians also differ, with \texttt{NEXONN} yielding $0.13^{+2.05}_{-2.14}$ compared with $-0.38^{+1.51}_{-1.80}$ from \texttt{PyMultiNest}. Although both distributions remain broad, these differences indicate that the two methods do not assign identical probability densities across the allowed cloud-pressure range.} In contrast, the cloud-fraction parameter, $\phi_{\mathrm{cloud}}$, shows the strongest posterior agreement among the eight MIRI/LRS parameters. \texttt{NEXONN} yields $0.55^{+0.33}_{-0.36}$ compared with $0.44^{+0.35}_{-0.31}$ from \texttt{PyMultiNest}, with $\mathrm{D}_{\mathrm{KL}}=0.06$, $\mathrm{D}_{\mathrm{JS}}=0.02$, \tr{an SWD of $0.30$}, and an overlap of $89.02\%$. The contrast among the aerosol parameters illustrates that the quality of posterior reproduction is parameter dependent even within the same retrieval and atmospheric model.

Figure~\ref{fig:miri spectrum} compares the transmission spectra corresponding to the \texttt{PyMultiNest} and \texttt{NEXONN} retrievals. Both median spectra reproduce the observed MIRI/LRS transmission spectrum and closely follow one another across the molecular features associated with $\mathrm{H_2O}$ and $\mathrm{SO_2}$. This agreement in spectral space demonstrates that the atmospheric solutions inferred by the two approaches provide comparably good descriptions of the observed spectrum. As in the simulated experiment, however, spectral agreement alone does not imply identical posterior distributions because atmospheric degeneracies can allow different combinations of parameters to produce similar spectra. \tr{In particular, the limited wavelength coverage and the interplay among molecular abundances, atmospheric temperature, reference pressure, and aerosol properties can permit different atmospheric parameter combinations to yield similar transmission spectra.} The statistical comparisons in Table~\ref{tab:MIRI_noise_comparison} therefore provide an essential complementary assessment of the inferred atmospheric constraints.

The corresponding marginalized and joint posterior distributions are compared in Figure~\ref{fig:miri_corner}. The figure visually confirms the parameter-dependent behavior quantified in Table~\ref{tab:MIRI_noise_comparison}: parameters such as $\mathrm{T_{iso}}$ and $\phi_{\mathrm{cloud}}$ show relatively strong posterior agreement, whereas the molecular abundances and several aerosol parameters show more noticeable differences in posterior width or detailed probability-density structure. \tr{The differences are particularly apparent for the haze-amplitude parameter, where the \texttt{NEXONN} posterior underrepresents part of the higher-amplitude region sampled by \texttt{PyMultiNest}. The two-dimensional projections provide a qualitative comparison of the parameter correlations and degeneracies, although the degree of correspondence varies among parameter pairs.} Because the statistical measures reported in Table~\ref{tab:MIRI_noise_comparison} are calculated from one-dimensional marginalized posteriors, Figure~\ref{fig:miri_corner} provides a complementary qualitative assessment of the multidimensional posterior structure. \tr{However, visual similarity between selected two-dimensional projections does not establish agreement of the complete eight-dimensional joint posterior, and the marginalized comparison metrics should not be interpreted as a quantitative validation of the full parameter-correlation structure.}

\tr{Taken together, the MIRI/LRS retrieval provides a complementary real-observation benchmark of \texttt{NEXONN} beyond the controlled simulated-spectrum experiment. Posterior overlaps range from $70.68\%$ to $89.02\%$, indicating substantial but parameter-dependent agreement with \texttt{PyMultiNest}. The logarithmic reference pressure, isothermal temperature, and cloud fraction show relatively strong agreement, whereas the larger KL divergences for $\mathrm{H_2O}$ and $\mathrm{SO_2}$ reveal differences in the detailed marginalized posterior densities despite broadly consistent central abundance estimates. The haze-amplitude and cloud-pressure posteriors show additional differences in their locations, widths, or probability-density structures, demonstrating that weakly constrained parameters can remain challenging for neural posterior approximation. Both retrieval approaches nevertheless provide similar spectral solutions under the adopted atmospheric model. These results establish that \texttt{NEXONN} can recover several principal atmospheric constraints obtained with \texttt{PyMultiNest} for the observed WASP-39~b MIRI/LRS spectrum, while also identifying parameter-specific discrepancies that require further investigation. Because the true atmospheric parameters are unknown and real observations may contain effects not represented in the training simulations, this comparison does not independently establish the accuracy or calibration of the inferred atmospheric posteriors. The statistical calibration of the \texttt{NEXONN} credible regions is therefore assessed separately using simulated spectra with known generating parameters through the coverage analysis discussed in Section~\ref{error propagation}.}

\begin{table*}
\centering
\caption{\tr{The Table below summarizes the comparative retrieval results of global wavelength coverage of JWST observations for WASP-39 b using \texttt{PyMultiNest} and \texttt{NEXONN} with \texttt{PandExo} noise obtained for the Hybrid Equilibrium chemistry model.}}
\label{tab:hybrid_comparison}

\renewcommand{\arraystretch}{1.2}

\begin{tabular}{lccccccc}
\hline\hline
Parameter & Training range$^{*}$ & \texttt{PyMultiNest} & \texttt{NEXONN} & KL Divergence & JS Divergence & Std. Wasserstein Distance & Overlap ($\%$) \\
\hline

$\log(\mathrm{P_{\mathrm{ref}}})$
& $\mathcal{U}(-6,2)$
& $-1.54^{+0.06}_{-0.03}$
& $-1.59^{+0.02}_{-0.07}$ 
& $1.98$ & $0.52$ & $1.08$ & $31.78$ \\

$\alpha_1$
& $\mathcal{U}(0.5,2)$
& $1.35^{+0.34}_{-0.42}$
& $1.66^{+0.09}_{-0.40}$ 
& $0.52$ & $0.17$ & $0.74$ & $54.85$ \\

$\alpha_2$
& $\mathcal{U}(0.5,2)$
& $1.47^{+0.28}_{-0.32}$
& $1.73^{+0.07}_{-0.35}$ 
& $0.53$ & $0.18$ & $0.80$ & $54.09$ \\

$\log(\mathrm{P_1})$
& $\mathcal{U}(-6,-4)$
& $-4.85^{+0.73}_{-0.93}$
& $-4.35^{+0.32}_{-1.10}$ 
& $0.24$ & $0.08$ & $0.51$ & $68.32$ \\

$\log(\mathrm{P_2)}$
& $\mathcal{U}(-2,-1)$
& $-1.93^{+1.01}_{-1.56}$
& $-1.45^{+0.79}_{-1.82}$ 
& $0.06$ & $0.02$ & $0.31$ & $84.94$ \\

$\log(\mathrm{P_3)}$
& $\mathcal{U}(1,2)$
& $1.02^{+0.47}_{-0.47}$
& $1.48^{+0.12}_{-0.20}$ 
& $0.80$ & $0.28$ & $0.96$ & $43.50$ \\

$\log(\mathrm{MgSiO_3})$
& $\mathcal{U}(-12,-5)$
& $-9.42^{+3.71}_{-4.87}$
& $-8.45^{+1.43}_{-3.89}$ 
& $0.11$ & $0.04$ & $0.28$ & $78.85$ \\

$\log(\mathrm{ZnS})$
& $\mathcal{U}(-7,-1)$
& $-2.54^{+0.74}_{-0.85}$
& $-1.81^{+0.37}_{-1.02}$ 
& $0.34$ & $0.12$ & $0.78$ & $64.10$ \\

$\log(\mathrm{Na})$
& $\mathcal{U}(-10,-4)$
& $-7.97^{+0.44}_{-0.50}$
& $-7.59^{+0.35}_{-0.33}$ 
& $0.24$ & $0.09$ & $0.82$ & $69.77$ \\

$\log(\mathrm{K})$
& $\mathcal{U}(-10,-5)$
& $-8.74^{+0.15}_{-0.16}$
& $-8.72^{+0.14}_{-0.20}$ 
& $0.01$ & $0.00$ & $0.10$ & $93.06$ \\

$\log(\mathrm{SO_2})$
& $\mathcal{U}(-8,-3)$
& $-5.80^{+0.11}_{-0.13}$
& $-5.68^{+0.05}_{-0.06}$ 
& $0.69$ & $0.25$ & $0.88$ & $48.48$ \\

$\mathrm{C/O}$
& $\mathcal{U}(0.3,1.2)$
& $0.80^{+0.03}_{-0.01}$
& $0.82^{+0.01}_{-0.02}$ 
& $0.25$ & $0.09$ & $0.53$ & $69.02$ \\

$[\mathrm{M/H}]$
& $\mathcal{U}(0,2)$
& $1.19^{+0.05}_{-0.04}$
& $1.13^{+0.04}_{-0.04}$ 
& $1.29$ & $0.30$ & $1.19$ & $47.45$ \\

$\mathrm{hc}$
& $\mathcal{U}(0.5,0.8)$
& $0.78^{+0.01}_{-0.01}$
& $0.77^{+0.01}_{-0.01}$ 
& $0.62$ & $0.19$ & $1.07$ & $55.83$ \\

$\mathrm{T_0}$
& $\mathcal{U}(600,1500)$
& $969.58^{+14.50}_{-8.54}$
& $978.04^{+8.56}_{-8.68}$ 
& $0.26$ & $0.10$ & $0.63$ & $68.02$ \\

$\mathrm{log(O/H)}^{**}$
& --
& $1.01^{+0.04}_{-0.04}$
& $0.95^{+0.03}_{-0.04}$
& -- & -- & -- & -- \\

$\mathrm{log(C/H)}^{**}$
& --
& $1.36^{+0.06}_{-0.04}$
& $1.31^{+0.04}_{-0.05}$
& -- & -- & -- & -- \\

$\mathrm{log(S/H)}^{**}$
& --
& $1.11^{+0.04}_{-0.03}$
& $1.05^{+0.03}_{-0.04}$
& -- & -- & -- & -- \\

\hline\hline
\end{tabular}

\begin{flushleft}
{\footnotesize $^{*}$ The training range represents the parameter ranges over which the \texttt{NEXONN} model was trained and the same range was used as the prior for the \texttt{PyMultiNest} retrieval. Here $\mathrm{hc}$ is the uniform cloud parameter and [M/H] is the global metallicity.}\\
{\footnotesize $^{**}$ These are the calculated elemental ratios and are not retrieved directly from the ML model. The elemental ratios tabulated here are obtained at $\mathrm{log(P_{ref}) = -2}$ bar.}
\end{flushleft}

\end{table*}

\begin{figure*}[ht]
    \centering

    \begin{minipage}{0.46\textwidth}
        \centering
        \includegraphics[width=\linewidth]{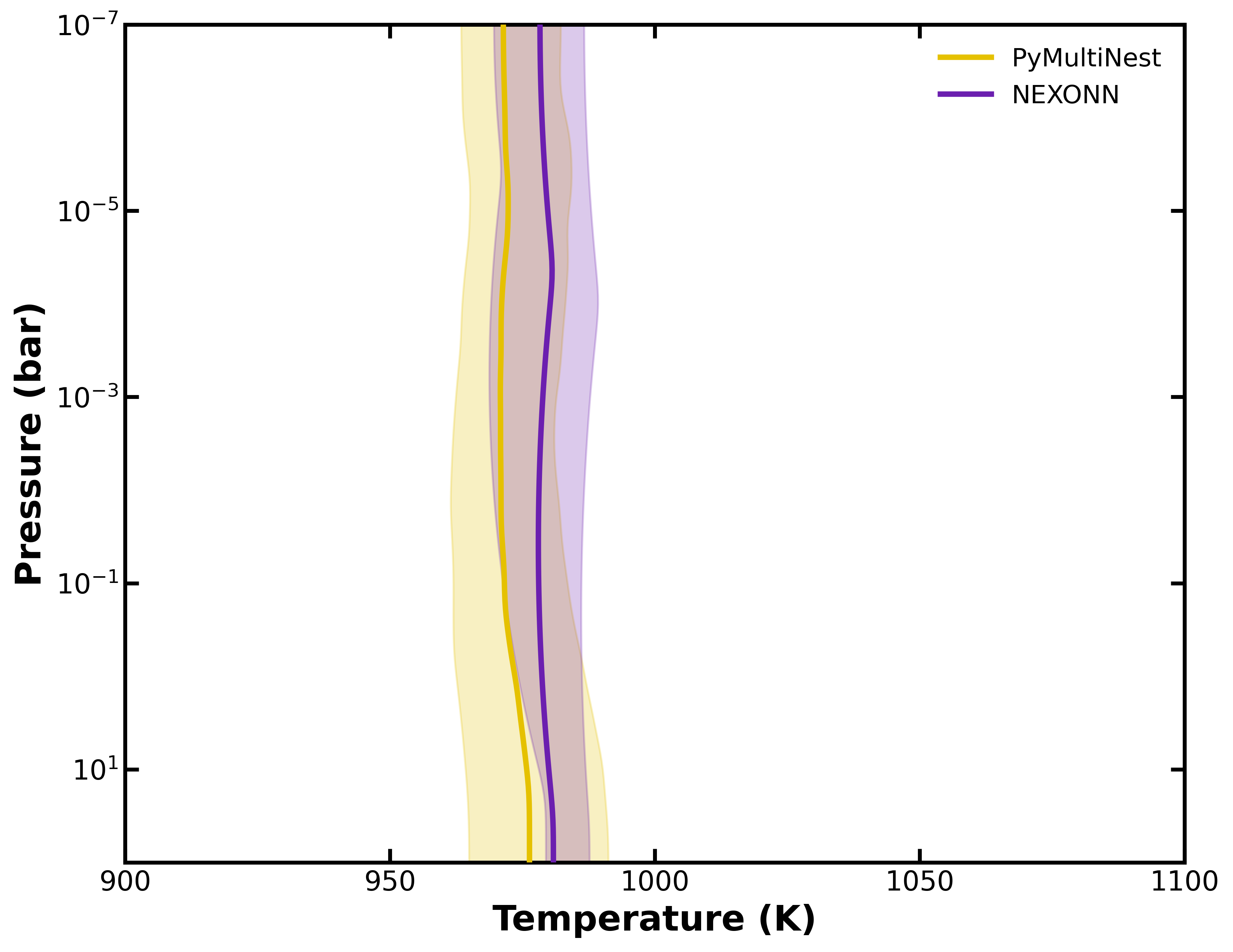}
        \caption{Retrieved pressure--temperature (P--T) profiles for the full JWST transmission spectrum of WASP-39~b using the Hybrid Equilibrium Chemistry model. The \texttt{PyMultiNest} and \texttt{NEXONN} results are shown in yellow and purple, respectively.}
        \label{fig:pt profile}
    \end{minipage}
    \hfill
    \begin{minipage}{0.52\textwidth}
        \centering
        \includegraphics[width=\linewidth]{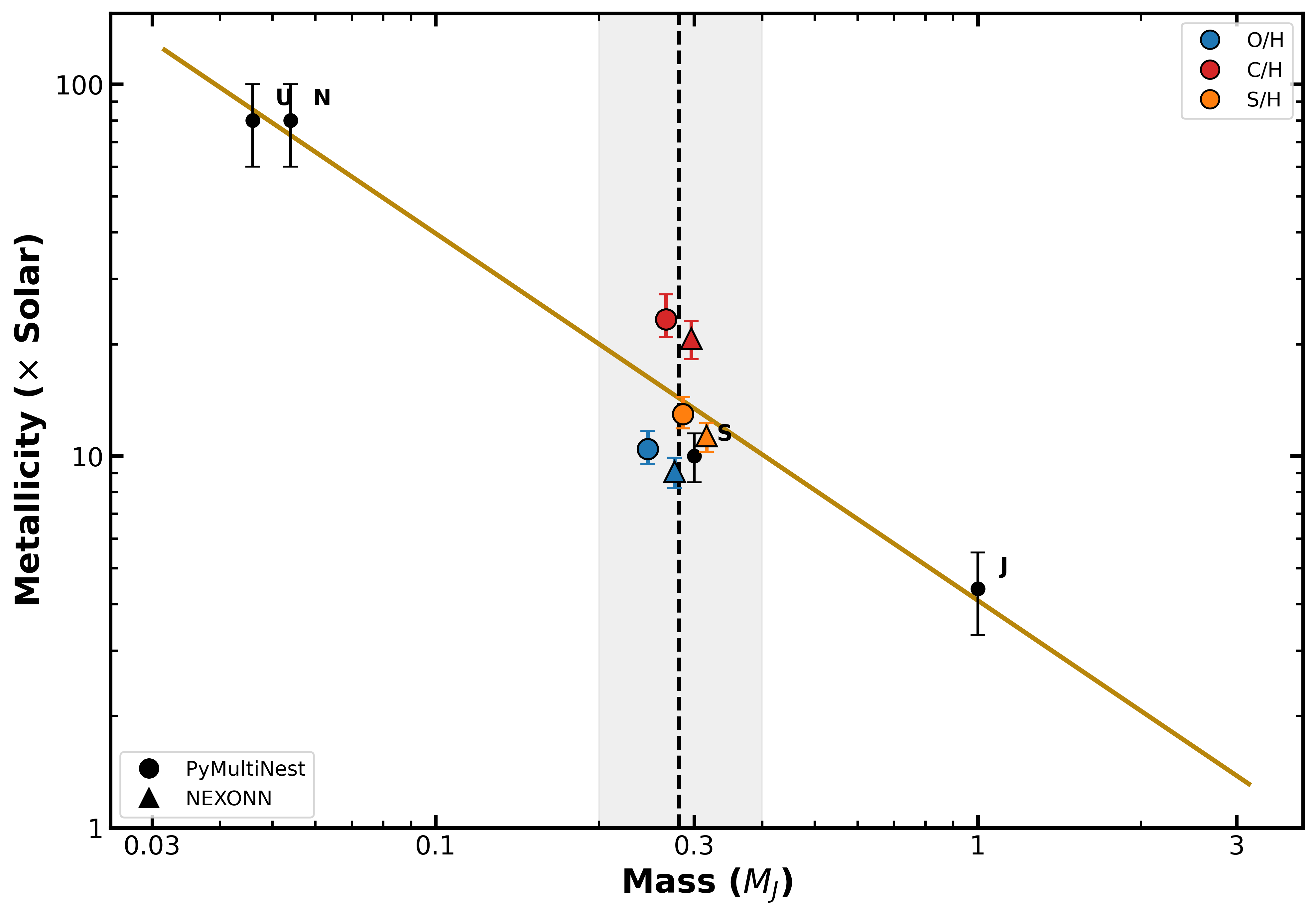}
        \caption{Mass--metallicity relation showing the atmospheric metallicity inferred for WASP-39~b using \texttt{PyMultiNest} and \texttt{NEXONN}.}
        \label{fig:metallicty}
    \end{minipage}

\end{figure*}

\subsection{\textbf{Applying \texttt{NEXONN} to the Full JWST Wavelength Coverage of WASP-39~b}}

To demonstrate the performance of \texttt{NEXONN} in a broad-wavelength retrieval, we use the combined JWST transmission spectrum of WASP-39~b spanning $0.6$--$12\,\mu\mathrm{m}$. The dataset comprises observations from multiple JWST instruments covering complementary wavelength regimes. We include the NIRISS/SOSS observations spanning $0.6$--$2.8\,\mu\mathrm{m}$ from \citet{Feinstein2023}, reduced using the \texttt{Supreme-SPOON} pipeline. We also incorporate the NIRSpec/PRISM observations from \citet{Rustamkulov2023}, which provide wavelength coverage from $0.5$ to $5.5\,\mu\mathrm{m}$ and were reduced using the \texttt{FIREFLy} pipeline. In this work, however, we use only the $2.0$--$5.5\,\mu\mathrm{m}$ region because the shorter-wavelength channels were reported to be affected by saturation \citep{Rustamkulov2023}. Finally, we include the MIRI/LRS observations spanning $5$--$12\,\mu\mathrm{m}$ from \citet{Powell2024}, reduced using the \texttt{Eureka!} pipeline. Together, these datasets provide continuous broad-wavelength coverage from the optical to the mid-infrared and probe a wide range of molecular and aerosol signatures.

The retrieval analysis adopts the best-fit atmospheric model identified by \citet{Deka_2025}, corresponding to the model with the lowest reduced $\chi^2$, namely the Hybrid Equilibrium Chemistry model. We benchmark the \texttt{NEXONN} results against the corresponding \texttt{PyMultiNest} retrieval. The forward model employs the Hybrid Equilibrium Chemistry framework, the Madhusudhan--Seager pressure--temperature (P--T) parameterization, and a uniform cloud and aerosol prescription. Within the Hybrid Equilibrium Chemistry framework, the abundances of $\mathrm{SO_2}$, $\mathrm{Na}$, and $\mathrm{K}$ are treated as free parameters and are therefore not constrained by the equilibrium-chemistry grid \texttt{NEXOCHEM}. In contrast, the abundances of $\mathrm{H_2O}$, $\mathrm{CO_2}$, $\mathrm{CO}$, and $\mathrm{H_2S}$ are determined under chemical-equilibrium constraints. The retrieved parameter constraints and their quantitative comparison with the corresponding \texttt{PyMultiNest} results are summarized in Table~\ref{tab:hybrid_comparison}.

The full-JWST retrieval represents a substantially more demanding test of \texttt{NEXONN} than the MIRI/LRS experiments because the atmospheric model contains 15 retrieved parameters describing the reference pressure, P--T structure, chemical composition, metallicity, C/O ratio, and clouds and aerosols. The statistical quantities in Table~\ref{tab:hybrid_comparison} compare the one-dimensional marginalized posterior distributions parameter by parameter, while Figure~\ref{fig:hybrid corner plot} provides a complementary view of the multidimensional posterior structure. \tr{The increased dimensionality introduces additional parameter correlations and degeneracies, making accurate posterior approximation more challenging. However, dimensionality alone cannot explain the observed differences between the retrieval methods, because observational information content, posterior geometry, and the relationship between the real observations and the simulated training distribution may also influence the results.} The comparison therefore tests not only whether \texttt{NEXONN} recovers similar central atmospheric constraints, but also how closely it reproduces the widths and probability-density structures of the corresponding \texttt{PyMultiNest} posteriors.

The transmission spectra corresponding to the retrieved median parameter values are shown in Figure~\ref{fig:hybrid spectrum}. Across the full wavelength range, the spectrum generated from the \texttt{NEXONN} retrieval closely follows the \texttt{PyMultiNest} result, with residuals of order $10^{-3}$. \tr{This close correspondence indicates that both methods identify atmospheric parameter combinations that reproduce similar transmission spectra under the adopted forward model.} As in the MIRI/LRS experiment, however, agreement in spectral space does not necessarily imply identical marginalized posterior distributions because correlated combinations of atmospheric parameters can produce similar spectra. \tr{In particular, the additional wavelength coverage can constrain combinations of molecular abundances, temperature structure, and aerosol properties without uniquely determining every individual model parameter.} The parameter-by-parameter statistical assessment in Table~\ref{tab:hybrid_comparison} therefore provides an important complementary test.

The principal bulk-composition parameters show \tr{different degrees of agreement between their central estimates and marginalized posterior distributions}. \texttt{NEXONN} retrieves a C/O ratio of $0.82^{+0.01}_{-0.02}$ compared with $0.80^{+0.03}_{-0.01}$ from \texttt{PyMultiNest}. \tr{The corresponding statistics are $\mathrm{D}_{\mathrm{KL}}=0.25$, $\mathrm{D}_{\mathrm{JS}}=0.09$, an SWD of $0.53$, and a posterior overlap of $69.02\%$. Thus, the central C/O estimates are similar, although the posterior overlap and distance statistics indicate differences in the marginalized probability distributions. For the global metallicity, \texttt{NEXONN} yields $\mathrm{[M/H]}=1.13^{+0.04}_{-0.04}$ compared with $1.19^{+0.05}_{-0.04}$ from \texttt{PyMultiNest}. Despite a difference of only $0.06$ dex between their medians, the metallicity posterior shows lower agreement, with $\mathrm{D}_{\mathrm{KL}}=1.29$, $\mathrm{D}_{\mathrm{JS}}=0.30$, an SWD of $1.19$, and an overlap of $47.45\%$. The two methods therefore infer broadly similar metallicity scales but assign appreciably different posterior probabilities within the corresponding parameter region. This distinction is important because agreement in the inferred atmospheric enrichment does not establish equivalent uncertainties or posterior calibration.}

The free chemical abundances show similarly parameter-dependent behavior. \tr{The strongest agreement is obtained for $\log(\mathrm{K})$, for which \texttt{NEXONN} gives $-8.72^{+0.14}_{-0.20}$ compared with $-8.74^{+0.15}_{-0.16}$ from \texttt{PyMultiNest}. All four statistical diagnostics indicate close posterior agreement, with $\mathrm{D}_{\mathrm{KL}}=0.01$, $\mathrm{D}_{\mathrm{JS}}=0.00$ at the reported precision, an SWD of $0.10$, and a posterior overlap of $93.06\%$. The Na abundance shows intermediate agreement, with a posterior overlap of $69.77\%$, $\mathrm{D}_{\mathrm{KL}}=0.24$, $\mathrm{D}_{\mathrm{JS}}=0.09$, and an SWD of $0.82$. In contrast, $\log(\mathrm{SO_2})$ yields median abundances of $-5.68^{+0.05}_{-0.06}$ from \texttt{NEXONN} and $-5.80^{+0.11}_{-0.13}$ from \texttt{PyMultiNest}, but a lower posterior overlap of $48.48\%$, together with $\mathrm{D}_{\mathrm{KL}}=0.69$, $\mathrm{D}_{\mathrm{JS}}=0.25$, and an SWD of $0.88$. The $0.12$ dex difference between the median $\mathrm{SO_2}$ abundances is accompanied by appreciable differences in posterior width and location. In particular, the \texttt{NEXONN} marginalized posterior is more concentrated than the corresponding \texttt{PyMultiNest} distribution according to the reported credible intervals. Thus, the chemical retrieval illustrates three distinct regimes: close posterior agreement for K, intermediate agreement for Na, and substantial posterior discrepancies for $\mathrm{SO_2}$ despite relatively similar central abundance estimates.}

The aerosol parameters also show parameter-dependent agreement. \tr{For $\log(\mathrm{MgSiO_3})$, the posterior overlap is $78.85\%$, with $\mathrm{D}_{\mathrm{KL}}=0.11$, $\mathrm{D}_{\mathrm{JS}}=0.04$, and an SWD of $0.28$. The $\log(\mathrm{ZnS})$ posterior has an overlap of $64.10\%$, with $\mathrm{D}_{\mathrm{KL}}=0.34$, $\mathrm{D}_{\mathrm{JS}}=0.12$, and an SWD of $0.78$. The uniform-cloud parameter, $\mathrm{h_{\mathrm{c}}}$, yields nearly identical central values, $0.77^{+0.01}_{-0.01}$ and $0.78^{+0.01}_{-0.01}$ for \texttt{NEXONN} and \texttt{PyMultiNest}, respectively. Nevertheless, its posterior overlap is only $55.83\%$, with $\mathrm{D}_{\mathrm{KL}}=0.62$, $\mathrm{D}_{\mathrm{JS}}=0.19$, and an SWD of $1.07$. The relatively low overlap and large standardized distance, despite the small difference between the median cloud parameters, indicate that the detailed posterior distributions are not reproduced equally well.} This again illustrates that proximity of the median values does not necessarily imply identical posterior widths or detailed probability-density structures.

The reference pressure provides the clearest example of this distinction. \texttt{NEXONN} retrieves $\log(\mathrm{P_{\mathrm{ref}}})=-1.59^{+0.02}_{-0.07}$ compared with $-1.54^{+0.06}_{-0.03}$ from \texttt{PyMultiNest}. \tr{Although the central values differ by only $0.05$ dex, the posterior overlap is $31.78\%$, the lowest among the 15 retrieved parameters. The corresponding $\mathrm{D}_{\mathrm{KL}}=1.98$ and $\mathrm{D}_{\mathrm{JS}}=0.52$ are the largest divergence values reported in Table~\ref{tab:hybrid_comparison}, while the SWD is $1.08$. The reported $68\%$ credible intervals also show limited overlap, with \texttt{NEXONN} spanning approximately $[-1.66,-1.57]$ and \texttt{PyMultiNest} spanning $[-1.57,-1.48]$. These results reveal greater differences in the concentration and distribution of posterior probability despite the proximity of the medians.} The reference-pressure result therefore demonstrates particularly clearly why comparison of median values or \tr{a single posterior-distance statistic} alone is insufficient for validating a posterior estimator. \tr{Because the reference pressure is correlated with the atmospheric opacity, composition, and thermal structure in transmission-spectrum retrievals, differences in its posterior distribution may also be associated with differences in the inferred parameter correlations. However, the one-dimensional statistics alone cannot establish the origin of these discrepancies.}

The parameters describing the Madhusudhan--Seager P--T profile show varying degrees of marginalized posterior agreement. \tr{The overlap is $54.85\%$ for $\alpha_1$, $54.09\%$ for $\alpha_2$, $68.32\%$ for $\log(\mathrm{P_1})$, $84.94\%$ for $\log(\mathrm{P_2)}$, and $43.50\%$ for $\log(\mathrm{P_3})$. The corresponding JS divergences are $0.17$, $0.18$, $0.08$, $0.02$, and $0.28$, respectively. The associated SWD values are $0.74$, $0.80$, $0.51$, $0.31$, and $0.96$.} Thus, $\log(\mathrm{P_2})$ is reproduced particularly well, whereas $\log(\mathrm{P_3})$ and the two $\alpha$ parameters show more noticeable differences in their marginalized posterior structures. The temperature normalization, $\mathrm{T_0}$, is broadly consistent between the two methods, with \texttt{NEXONN} yielding \tr{$978.04^{+8.56}_{-8.68}\,\mathrm{K}$} compared with $969.58^{+14.50}_{-8.54}\,\mathrm{K}$ from \tr{\texttt{PyMultiNest}. The corresponding posterior overlap is $68.02\%$, with $\mathrm{D}_{\mathrm{KL}}=0.26$, $\mathrm{D}_{\mathrm{JS}}=0.10$, and an SWD of $0.63$. The different posterior constraints on the profile parameters indicate that the TcNF does not reproduce every aspect of the marginalized thermal-structure posterior, even when the characteristic temperature scales are similar.}

Figure~\ref{fig:pt profile} compares the P--T profiles resulting from these correlated profile parameters. Despite the parameter-dependent differences in their individual marginalized posteriors, the resulting atmospheric thermal structures are broadly consistent between the two retrieval methods. This illustrates an important property of the P--T parameterization: different combinations of correlated profile parameters can produce similar temperature structures over the pressure range probed by the transmission spectrum. Consequently, differences among individual P--T parameters should be interpreted together with the resulting P--T profile rather than in isolation. \tr{The agreement between the plotted thermal structures therefore provides a complementary physical comparison, but does not by itself establish that the two methods recover identical posterior uncertainties in temperature as a function of pressure. Such a conclusion would require a direct comparison of posterior predictive P--T distributions obtained from the full joint posterior samples.}

The derived elemental abundance ratios provide an additional consistency check on the broader chemical solution. \tr{These quantities are calculated from the retrieved atmospheric composition at a reference pressure of $10^{-2}\,\mathrm{bar}$ and are not directly inferred as independent parameters by \texttt{NEXONN}.} We obtain $\log(\mathrm{O/H})=0.95^{+0.03}_{-0.04}$, $\log(\mathrm{C/H})=1.31^{+0.04}_{-0.05}$, and $\log(\mathrm{S/H})=1.05^{+0.03}_{-0.04}$ from the \texttt{NEXONN} retrieval, compared with $1.01^{+0.04}_{-0.04}$, $1.36^{+0.06}_{-0.04}$, and $1.11^{+0.04}_{-0.03}$, respectively, from the \texttt{PyMultiNest} retrieval. \tr{The differences between the corresponding central estimates are $0.06$, $0.05$, and $0.06$ dex for O/H, C/H, and S/H, respectively. These results indicate broadly similar inferred elemental-enrichment patterns under the adopted chemical model. However, because the elemental ratios are derived from the retrieved atmospheric composition, their apparent agreement does not constitute an independent validation of the underlying posterior distributions.}

The mass--metallicity relation is shown in Figure~\ref{fig:metallicty}. The inferred atmospheric metallicity places WASP-39~b in an enriched regime relative to Solar System giant planets of comparable mass, with potential implications for its formation and evolutionary history \citep{Madhusudhan_2012}. In the present context, the important methodological result is that both inference approaches place WASP-39~b in a similar region of the mass--metallicity relation, even though the statistical comparison shows that the detailed $\mathrm{[M/H]}$ posterior densities are not identical. \tr{Consequently, the broad interpretation of an enriched atmosphere is shared by the two retrievals, whereas the precise metallicity constraint and its uncertainty remain dependent on the inferred posterior distribution and the adopted atmospheric model.}

Figure~\ref{fig:hybrid corner plot} compares the joint and marginalized posterior distributions obtained with \texttt{PyMultiNest} and \texttt{NEXONN}. The figure visually reinforces the parameter-dependent behavior quantified in Table~\ref{tab:hybrid_comparison}. Some parameters, most notably $\log(\mathrm{K})$, show closely matching marginalized distributions, whereas parameters such as $\log(\mathrm{P_{\mathrm{ref}}})$, $\log(\mathrm{SO_2})$, $\mathrm{[M/H]}$, and several P--T parameters show more noticeable differences in posterior width or detailed shape. \tr{The discrepancies in these comparatively concentrated posteriors are particularly relevant because relatively small changes in posterior location or width can produce substantial changes in their overlap and standardized distance. The two-dimensional projections provide a qualitative assessment of the inferred parameter correlations and degeneracies, but the degree of correspondence is not uniform across parameter pairs.} Because the statistical measures in Table~\ref{tab:hybrid_comparison} are calculated from one-dimensional marginalized posteriors, Figure~\ref{fig:hybrid corner plot} should be viewed as a complementary visual assessment of the multidimensional posterior structure rather than as a quantitative validation of the full 15-dimensional joint distribution. \tr{A rigorous assessment of joint-posterior fidelity would additionally require multivariate diagnostics capable of testing the correlations and dependence structures represented by the TcNF.}

\tr{Taken together, the full-JWST experiment demonstrates that the performance of \texttt{NEXONN} cannot be characterized by a single measure of posterior agreement. The framework recovers broadly similar central constraints for several physically important quantities, including C/O, metallicity, selected chemical abundances, and the characteristic thermal structure, while the statistical assessment reveals substantial parameter-to-parameter variation in the marginalized posterior distributions. Posterior overlaps range from $31.78\%$ to $93.06\%$, with the lowest and highest overlaps corresponding to $\log(\mathrm{P_{\mathrm{ref}}})$ and $\log(\mathrm{K})$, respectively. Although $\log(\mathrm{K})$ shows close agreement across all four statistical measures, the reference pressure, metallicity, $\mathrm{SO_2}$ abundance, and several P--T parameters show appreciable differences in posterior location, width, or probability-density structure. These discrepancies are particularly important because agreement in the reconstructed transmission spectrum or retrieved median parameters does not guarantee accurate approximation of the underlying Bayesian posterior. The full-JWST analysis therefore demonstrates the potential of \texttt{NEXONN} for rapid atmospheric inference while identifying important limitations in posterior fidelity for this higher-dimensional retrieval. The possible contributions of atmospheric degeneracies, finite training-set coverage, posterior complexity, and differences between simulated training spectra and real observations are discussed further in Section~\ref{sec:discussion}. These mechanisms remain possible explanations rather than established causes of the observed discrepancies.}

\begin{figure}[ht]
    \centering
    \includegraphics[width=0.9\linewidth]{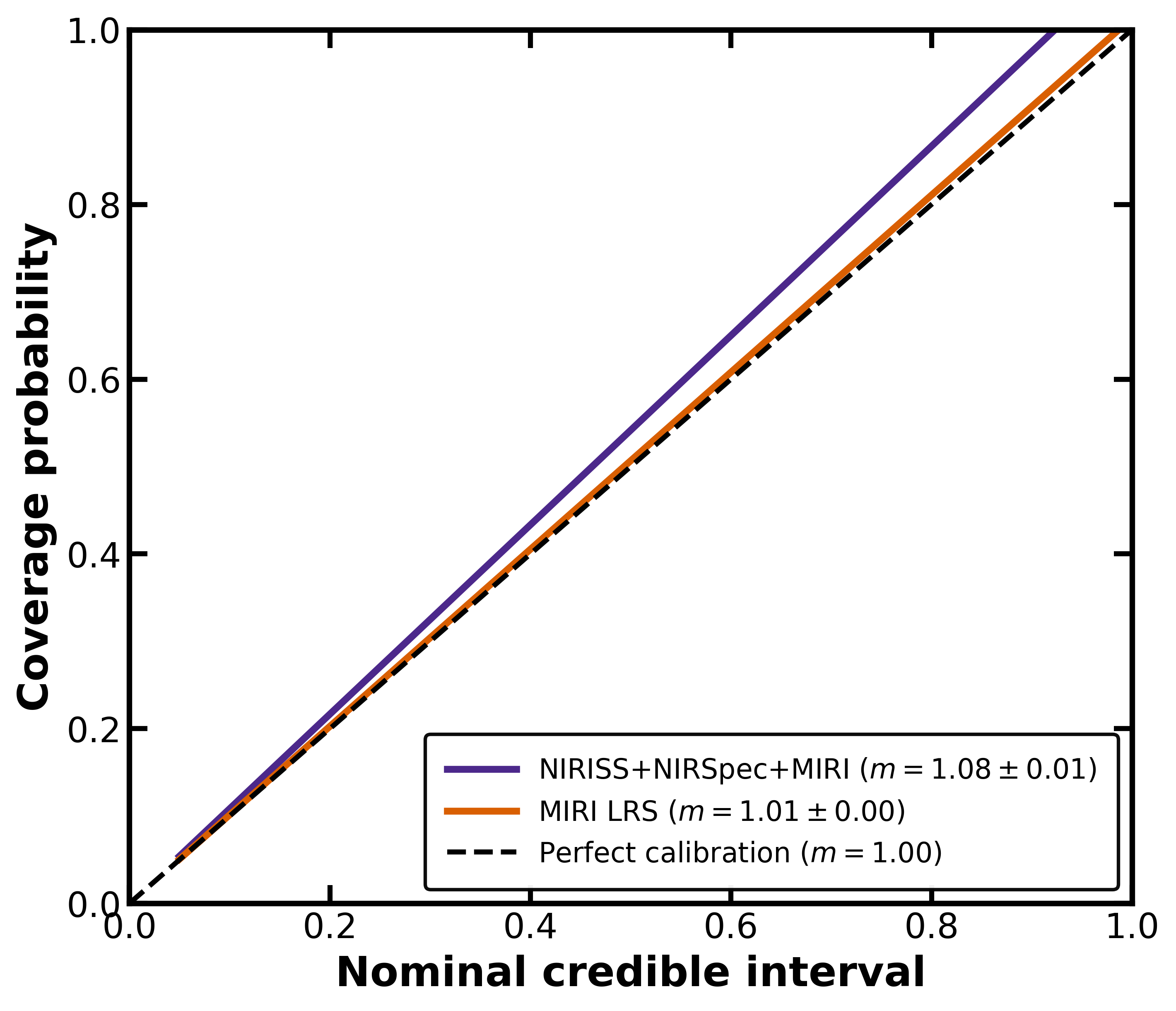}
    \caption{Coverage analysis of the posterior distributions, $P(\theta\mid\mathbf{x})$, inferred with \texttt{NEXONN} for the JWST MIRI/LRS retrieval and the full JWST retrieval combining NIRISS/SOSS, NIRSpec/PRISM, and MIRI/LRS. Each curve represents the average empirical coverage across all retrieved parameters for 500 test spectra. The curves are linearly fitted and compared with the ideal one-to-one calibration relation. The close agreement between the empirical and nominal credible levels indicates that the inferred posterior uncertainties are well calibrated.}
    \label{coverage plot}
\end{figure}

\vspace{1cm}
\section{\textbf{Discussion}}
\label{sec:discussion}

The retrieval experiments presented above show that \texttt{NEXONN} can recover the principal atmospheric constraints obtained with conventional Bayesian sampling while reducing the cost of posterior inference after training. At the same time, the statistical comparisons demonstrate that agreement between neural and Bayesian inference is not uniform across all parameters and cannot be assessed from median values alone. In this section, we interpret these results in the context of posterior approximation, atmospheric degeneracies, statistical calibration, and the computational advantages and limitations of amortized inference.

\subsection{\textbf{Comparison of \texttt{PyMultiNest} and \texttt{NEXONN} Retrievals}}

The three retrieval experiments provide complementary tests of increasing complexity. The simulated MIRI/LRS experiment provides the most controlled validation because the generating atmospheric parameters are known and the inference problem is restricted to eight parameters. The observed MIRI/LRS retrieval then tests whether the trained neural posterior estimator remains consistent with conventional Bayesian inference when applied to real JWST data. Finally, the full-wavelength retrieval extends the comparison to a 15-dimensional atmospheric model containing a more complex thermal structure, equilibrium and free-chemistry parameters, and aerosol properties. Considered together, these experiments therefore probe the transition from controlled validation to increasingly realistic and degenerate atmospheric inference.

A central result of these comparisons is that similarity of retrieved median values does not necessarily imply similarity of the corresponding posterior distributions. This distinction becomes increasingly important as the retrieval dimensionality and parameter correlations increase. The KL and JS divergences, \tr{SWD}, and posterior overlap probe complementary aspects of the marginalized posterior distributions and therefore reveal differences that would be hidden by a comparison of best-fit or median values alone. In particular, two posteriors can occupy nearby regions of parameter space while differing in their widths, asymmetries, tails, or concentration of probability density. Conversely, a high posterior overlap for a broad distribution does not imply that the underlying atmospheric parameter is tightly constrained by the data \tr{and vice versa}.

The simulated-spectrum experiment illustrates the latter point particularly clearly. Several cloud and haze parameters remain weakly constrained by the adopted MIRI/LRS spectrum, yet \texttt{NEXONN} reproduces the broad \texttt{PyMultiNest} posteriors with high fidelity. This behavior is important for a neural posterior estimator: successful inference should reproduce not only well-constrained parameters but also the absence of information when the data do not strongly constrain a parameter. The agreement in these weakly informative dimensions therefore indicates that \texttt{NEXONN} is learning aspects of the posterior uncertainty structure rather than merely mapping spectra to point estimates.

For the observed WASP-39~b retrievals, the posterior agreement becomes more parameter dependent. The MIRI/LRS experiment shows that the two approaches can recover similar physical and chemical solutions even when the detailed probability density differs for individual parameters. This is especially relevant for molecular abundances, where the characteristic abundance constraints may agree while the posterior-comparison metrics indicate differences in posterior shape or density. \tr{In particular, Figure~6 shows differences in the $\log(a)$ posterior, where \texttt{NEXONN} assigns less probability to the larger values favored by \texttt{PyMultiNest}, while the $\gamma$ and $\log(\mathrm{P_{\mathrm{cloud}}})$ posteriors are more tightly constrained by \texttt{NEXONN}.} The full-JWST retrieval strengthens this conclusion: several scientifically important quantities, including C/O, metallicity, elemental enrichment, and the atmospheric thermal structure, remain broadly consistent between the two inference approaches, while some individual marginalized posteriors show substantially lower overlap. \tr{In Figure~8, the differences are particularly noticeable for $\log(\mathrm{P_{\mathrm{ref}}})$, $\log(\mathrm{M/H})$, and the individual P--T profile parameters, highlighting that agreement in characteristic parameter values does not necessarily imply agreement in the full posterior distributions. Thus, for the relatively well-constrained WASP-39~b observations considered here, the two methods yield broadly consistent atmospheric interpretations despite noticeable differences in some marginalized posteriors. However, this agreement should not be generalized to less informative observations or more strongly degenerate retrieval problems, where such posterior differences could lead to substantially different atmospheric interpretations. A systematic assessment of \texttt{NEXONN} under these conditions is left for future work.}

The P--T retrieval provides an instructive example of the role of parameter correlations. Individual parameters of the Madhusudhan--Seager profile need not show identical marginalized posteriors for the two inference methods, because multiple correlated combinations of these parameters can produce similar temperature structures over the pressures probed by transmission spectroscopy. The agreement of the resulting P--T profiles therefore carries different information from agreement of the individual parameter marginals. More generally, atmospheric retrieval parameters should not always be interpreted independently when the forward model admits strong covariances or degeneracies. \tr{Nevertheless, agreement between the derived P--T profiles does not eliminate differences between the underlying parameter posteriors, which remain relevant when assessing the accuracy of the neural posterior approximation.}

\tr{The progression from the simulated experiment to the real JWST observations reveals an important limitation of the present approach. Although the simulated and observed MIRI/LRS retrievals involve the same eight parameters, \texttt{NEXONN} shows closer agreement with \texttt{PyMultiNest} for the simulated spectrum (Figure~4) than for the observed spectrum (Figure~6). This suggests that the differences cannot be attributed solely to increased retrieval dimensionality.}

\tr{A plausible explanation is a mismatch between the training distribution and real observations. \texttt{NEXONN} is trained on synthetic spectra generated using an idealized one-dimensional atmospheric forward model and the adopted \texttt{PandExo}-based noise prescription. In contrast, real exoplanet atmospheres are three-dimensional and may show spatially varying temperatures, chemical abundances, and cloud properties. Furthermore, real JWST observations may contain instrumental systematics and noise characteristics not fully represented by the training simulations.}

\tr{Consequently, even when the retrieved atmospheric parameters lie within the adopted training ranges, the observed spectra may differ from the distribution of spectra encountered during training. The application to real JWST observations may therefore constitute an out-of-distribution (OOD) inference problem, potentially contributing to the reduced posterior agreement with \texttt{PyMultiNest}.}

\tr{Additional factors, including finite training-set coverage, parameter degeneracies, and increasingly complex posterior correlations, may also affect the accuracy of the neural posterior approximation, particularly for the 15-dimensional full-JWST retrieval. However, the present experiments do not isolate the relative contributions of these effects or establish whether atmospheric model inadequacies or differences in the observational noise prescription are responsible for the observed discrepancies.}

\tr{These results highlight the importance of evaluating neural posterior estimators beyond the synthetic distributions used during training. While the present experiments demonstrate promising performance for WASP-39~b, agreement with conventional Bayesian retrievals on simulated spectra does not guarantee equivalent posterior agreement on real observations. Future work should therefore investigate the effects of training-set coverage, atmospheric model mismatch, realistic instrumental noise, and OOD behavior through controlled experiments, particularly before extending \texttt{NEXONN} to planets with lower-quality observations or substantially different atmospheric properties.}

\subsubsection{\textbf{Posterior Uncertainties and Calibration}}
\label{error propagation}

The use of \texttt{PandExo}-based noise realizations exposes
\texttt{NEXONN} during training to wavelength-dependent uncertainty
structures representative of the adopted JWST observing configurations.
This differs from training on spectra perturbed by a wavelength-independent
random-noise prescription and allows the observational information content
to vary across wavelength in a manner that reflects the adopted
instrumental setup. The present analysis does not, however, include
an ablation experiment comparing different noise prescriptions;
therefore, the posterior agreement reported here should be interpreted
as demonstrating the performance of \texttt{NEXONN} under the adopted
\texttt{PandExo}-based treatment rather than establishing that this
treatment itself is responsible for the agreement with \texttt{PyMultiNest}.
\tr{However, the adopted \texttt{PandExo}-based noise prescription
remains an approximation of the observational uncertainties and may
not fully capture instrumental systematics or correlated noise
present in real JWST observations. Investigating more realistic
noise prescriptions and their effects on posterior inference
will be an important direction for future work with
\texttt{NEXONN}.}

The widths and shapes of the \texttt{NEXONN} posteriors represent
uncertainty conditional on the adopted forward model, the training
distribution, and the observational noise prescription. Parameters
associated with distinctive spectral information can therefore yield
concentrated posteriors, whereas weakly constrained or strongly
degenerate parameters can remain broad. This behavior is important
for a posterior estimator because the absence of strong spectral
information should be represented as posterior uncertainty rather
than as artificially precise parameter estimates. However, these
uncertainties should not be interpreted as accounting for model
inadequacy or systematic errors outside the assumptions represented
in the training data and forward model.

Agreement with \texttt{PyMultiNest} and statistical calibration
address two distinct questions. The posterior-comparison metrics
quantify how closely \texttt{NEXONN} approximates a particular
Bayesian reference posterior, whereas coverage tests whether the
credible regions produced by \texttt{NEXONN} contain the known
parameters at the expected frequency over repeated realizations.
A neural posterior may therefore be statistically calibrated even
when its marginalized posterior differs in detail from the
\texttt{PyMultiNest} result for an individual spectrum. Conversely,
close agreement with a Bayesian posterior for a single observation
does not by itself demonstrate correct repeated-sampling calibration.

\tr{The coverage analysis in Figure~\ref{coverage plot} shows that
the \texttt{NEXONN} credible regions show good aggregate calibration
under the adopted simulation and noise assumptions. The model trained
for the JWST MIRI/LRS retrieval yields a fitted coverage slope of
$m=1.01\pm0.00$ with a coefficient of determination of $R^2=0.996$,
placing it very close to the ideal one-to-one calibration relation.
For the full JWST model spanning the combined NIRISS/SOSS,
NIRSpec/PRISM, and MIRI/LRS wavelength ranges, the fitted slope is
$m=1.08\pm0.01$ with $R^2=0.989$. The full-JWST model therefore
shows a mild tendency toward conservative (underconfident)
coverage, with empirical coverage exceeding the nominal credible
levels on average over the fitted range. This suggests that the
inferred credible regions are somewhat broader than required for
exact calibration under the adopted simulation assumptions.
The high $R^2$ values indicate that the empirical coverage is
well described by the fitted relations, although they do not
independently establish agreement with the ideal one-to-one
relation. Importantly, these coverage curves represent averages
across all retrieved parameters for 500 simulated test spectra.
They therefore characterize aggregate calibration under the
training simulation framework, rather than establishing exact
calibration for every individual parameter or for real JWST
observations. Parameter-resolved coverage tests and validation
under more realistic observational conditions would provide
useful extensions in future work.}

The slightly more conservative behavior of the full-JWST model
is qualitatively consistent with the increased complexity of the
15-dimensional inference problem, although the present analysis
does not establish a causal connection. Higher dimensionality,
stronger posterior correlations, and finite sampling of the
training distribution could all contribute to broader inferred
uncertainties. More extensive simulation-based calibration,
including parameter-resolved and potentially joint-posterior
diagnostics, will therefore be valuable as \texttt{NEXONN}
is extended to more complex atmospheric models.

\subsubsection{\textbf{Computational Cost}}
\vspace{-0.2cm}
The principal computational trade-off of \texttt{NEXONN} is the shift from repeated retrieval cost to an up-front training cost. For the full-JWST configuration considered here, training on $2\times10^{6}$ synthetic spectra in the 15-dimensional parameter space at a spectral resolution of $R=3000$ required approximately $36$ hours on an NVIDIA L4 GPU. Once trained, however, posterior inference for an individual spectrum required only $8.14$~s on the same GPU. The corresponding \texttt{PyMultiNest} retrieval required approximately $49$ hours on 60 CPU cores, or about 2915 CPU-hours. Because these calculations use fundamentally different algorithms and hardware architectures, these wall times should not be interpreted as a direct hardware-normalized speed-up factor. Instead, the relevant advantage is the different scaling with the number of retrieval targets.

A conventional Bayesian retrieval incurs a new sampling cost for every spectrum, whereas the dominant training cost of an amortized neural posterior estimator is paid once. For a set of spectra that can be analyzed using the same trained model and that remain within its applicable training domain, the marginal computational cost per additional target is therefore very small. This distinction is particularly important for large homogeneous datasets, repeated analyses, injection-and-recovery experiments, and population-level atmospheric studies, where many posterior evaluations may be required.

The current training cost nevertheless remains substantial and motivates further optimization of the neural architecture and training strategy. More efficient attention implementations and improved simulation and training strategies may reduce this cost in future versions of \texttt{NEXONN}. Such developments should be evaluated not only in terms of training speed but also in terms of posterior fidelity, statistical calibration, memory requirements, and robustness across the training domain.

The scalability of amortized inference must also be considered together with its domain dependence. A trained \texttt{NEXONN} model cannot automatically be assumed to generalize to arbitrary wavelength coverage, noise properties, atmospheric parameterizations, or planets lying outside the training distribution. Changes to the forward model or observational configuration may require additional training or retraining. Consequently, the greatest computational advantage is expected when the up-front cost can be amortized over many spectra sharing a sufficiently similar inference problem. This regime is particularly relevant to population-level atmospheric studies and comparative exoplanetology, where consistently analyzing large samples within a common physical framework is central to identifying population-level trends.

\vspace{1cm}
\section{\textbf{Conclusion}}
\label{sec:conclusion}

In this work, we introduced \texttt{NEXONN}, a machine-learning retrieval framework within \texttt{NEXOTRANS} that combines a \tr{transformer-based conditioner with normalizing flows} for amortized posterior estimation. We evaluated \texttt{NEXONN} through a sequence of increasingly complex retrieval experiments, benchmarking its posterior constraints against Bayesian retrievals performed with \texttt{PyMultiNest}. Together, these experiments assess the recovery of atmospheric constraints, the fidelity of the inferred posterior distributions, their statistical calibration, and the computational advantages of amortized inference. \tr{The comparison distinguishes agreement in retrieved atmospheric properties from agreement in posterior probability distributions, which is essential for assessing the reliability of neural posterior estimation.} The main conclusions are summarized below:

\begin{enumerate}
\renewcommand{\labelenumi}{(\roman{enumi})}

    \item We first validated \texttt{NEXONN} using a simulated JWST MIRI/LRS transmission spectrum spanning $5$--$12\,\mu\mathrm{m}$, with \texttt{PandExo}-based noise incorporated into the synthetic spectrum. Because the input atmospheric parameters are known, this experiment provides a controlled validation of the retrieval framework. \tr{\texttt{NEXONN} successfully recovers atmospheric solutions consistent with the simulated spectrum and reproduces the principal characteristics of the corresponding \texttt{PyMultiNest} marginalized posterior distributions. The mean posterior overlap across the eight retrieved parameters is $80.32\%$, reaching $94.92\%$ for $\log(a)$, with particularly strong agreement for the cloud and haze parameters. Although some differences remain for $\mathrm{\mathrm{T_{\mathrm{iso}}}}$ and $\log(\mathrm{SO_2})$, the overall results demonstrate substantial consistency between the two retrieval approaches. Importantly, \texttt{NEXONN} also reproduces the broad posterior distributions of weakly constrained parameters, demonstrating its ability to capture atmospheric uncertainties and parameter degeneracies rather than providing only point estimates.}

    \item We next applied \texttt{NEXONN} to the observed JWST MIRI/LRS transmission spectrum of WASP-39~b spanning $5$--$12\,\mu\mathrm{m}$. The retrieval recovers the principal atmospheric constraints obtained with \texttt{PyMultiNest}, \tr{with substantial agreement across the marginalized posterior distributions.} In particular, \texttt{NEXONN} retrieves $\log(\mathrm{H_2O})=-1.31^{+0.41}_{-0.50}$ and $\log(\mathrm{SO_2})=-5.60^{+0.74}_{-0.47}$, compared with the \texttt{PyMultiNest} constraints of $\log(\mathrm{H_2O})=-1.44^{+0.24}_{-0.45}$ and $\log(\mathrm{SO_2})=-5.71^{+0.37}_{-0.41}$, respectively. \tr{The corresponding posterior overlaps of $71.97\%$ for $\log(\mathrm{H_2O})$ and $79.41\%$ for $\log(\mathrm{SO_2})$ demonstrate substantial agreement in the inferred molecular abundances. The cloud and haze parameters also show broadly consistent posterior constraints, although some differences in their distributions remain, reflecting the parameter-dependent nature of posterior approximation.} \tr{Overall, this real-data experiment demonstrates that \texttt{NEXONN} successfully reproduces the principal molecular and atmospheric constraints inferred from JWST observations, while providing a probabilistic characterization of the associated uncertainties.} 
    
    \item We then extended \texttt{NEXONN} to the full JWST transmission spectrum of WASP-39~b, combining NIRISS/SOSS, NIRSpec/PRISM, and MIRI/LRS observations over approximately $0.6$--$12\,\mu\mathrm{m}$. Using the Hybrid Equilibrium Chemistry model, \texttt{NEXONN} yields a C/O ratio of $0.82^{+0.01}_{-0.02}$ and a global metallicity of $\mathrm{[M/H]}=1.13^{+0.04}_{-0.04}$, compared with the corresponding \texttt{PyMultiNest} constraints of $0.80^{+0.03}_{-0.01}$ and \tr{$\mathrm{[M/H]}=1.19^{+0.05}_{-0.04}$}, respectively. The derived logarithmic elemental abundance ratios, $\log(\mathrm{O/H})=0.95^{+0.03}_{-0.04}$, $\log(\mathrm{C/H})=1.31^{+0.04}_{-0.05}$, and $\log(\mathrm{S/H})=1.05^{+0.03}_{-0.04}$, are also consistent with those derived from the \texttt{PyMultiNest} retrieval. \tr{The retrieved P--T profiles also show broad agreement, demonstrating that both approaches recover similar atmospheric thermal structures despite differences in some of the individual P--T parameters. Although the detailed posterior distributions show parameter-dependent differences, including those associated with metallicity and reference pressure, the overall results demonstrate that \texttt{NEXONN} successfully recovers the principal chemical and thermal characteristics of WASP-39~b from the combined JWST observations. This agreement is particularly encouraging given the increased complexity of the 15-dimensional retrieval and highlights the potential of \texttt{NEXONN} for efficient atmospheric characterization using broad-wavelength JWST spectroscopy.}

    \item Across the simulated, MIRI/LRS, and full-JWST experiments, we quantified posterior agreement using complementary statistical measures, including KL divergence, JS divergence, \tr{standardized Wasserstein distance}, and posterior overlap. These diagnostics demonstrate that agreement in median retrieved values does not necessarily imply agreement of the full marginalized posterior distributions. Some parameters show close agreement across all statistical measures, whereas others occupy similar regions of parameter space but differ in posterior width, asymmetry, or probability-density structure. \tr{The discrepancies become particularly evident in the full-JWST experiment, where several scientifically important parameters show relatively similar central estimates but different posterior overlaps and divergence statistics. Moreover, the one-dimensional metrics assess marginalized distributions and cannot, by themselves, establish agreement of the full multidimensional posterior or its parameter correlations.} The statistical assessment therefore highlights the importance of evaluating neural posterior estimators using distribution-level diagnostics rather than central parameter estimates alone.

    \item Posterior fidelity with respect to \texttt{PyMultiNest} and statistical calibration were assessed as complementary but distinct properties. \tr{Simulation-based coverage tests indicate that the \texttt{NEXONN} credible regions show approximately nominal or mildly conservative aggregate coverage under the adopted forward-model and noise assumptions.} The MIRI/LRS model yields a coverage slope of $m=1.01\pm0.00$ with $\mathrm{R^2=0.996}$, while the full-JWST model yields $m=1.08\pm0.01$ with $\mathrm{R^2=0.989}$. The former is close to the ideal calibration relation, whereas the latter indicates mildly conservative credible regions for the higher-dimensional retrieval. Because these coverage curves are averaged across the retrieved parameters, these results establish \tr{approximate aggregate calibration on the simulated test population rather than calibrated uncertainties for every individual parameter or real observation. In particular, the mildly conservative aggregate coverage of the full-JWST model does not contradict the narrower \texttt{NEXONN} posteriors observed for some parameters in the WASP-39~b retrieval, because the two diagnostics evaluate different properties. Parameter-resolved coverage tests and more extensive validation of the joint posterior will be necessary to establish the reliability of the inferred uncertainties across individual atmospheric parameters and observational conditions.}

    \item The \texttt{PandExo}-based noise prescription incorporates wavelength-dependent observational uncertainty structures representative of the adopted JWST observing configurations into the synthetic training spectra. \texttt{NEXONN} therefore learns posterior inference under an instrument-dependent noise treatment rather than a wavelength-independent random-noise prescription. The results presented here demonstrate the performance of the framework under this adopted noise model; quantifying the specific impact of the noise prescription itself will require dedicated comparisons between alternative training-noise treatments. \tr{Furthermore, the adopted noise prescription does not necessarily capture all instrumental systematics, correlated observational errors, or departures of real planetary atmospheres from the assumptions of the training forward model. Such differences may affect the transfer of the trained posterior estimator to real observations and warrant further investigation.} 
    
    \item The \tr{transformer-conditioned normalizing flow architecture} enables \texttt{NEXONN} to \tr{learn complex, multidimensional posterior distributions and capture correlations and degeneracies among atmospheric parameters, providing probabilistic constraints rather than only point estimates.} This represents a substantial extension of the previous machine-learning implementation within \texttt{NEXOTRANS}, which employed a Stacking Regressor combining Random Forest, Gradient Boosting, and K-Nearest Neighbors. \tr{By extending the framework from deterministic parameter estimation to amortized neural posterior estimation, \texttt{NEXONN} provides a flexible and computationally efficient approach to atmospheric retrieval with uncertainty quantification.} \tr{Our comparisons of marginalized posterior distributions, together with qualitative assessments of parameter correlations, demonstrate the potential of this approach for increasingly complex atmospheric models. Future developments will focus on quantitative validation of multidimensional posterior structure and parameter-specific calibration, further strengthening the reliability and applicability of the framework.}

    \item Finally, \texttt{NEXONN} provides a substantial computational advantage through amortized inference. Training the full-JWST model on $2\times10^{6}$ synthetic spectra requires $\sim36$ hours on an NVIDIA L4 GPU. Once trained, posterior sampling for an individual spectrum requires only $8.14$~s \tr{on the same GPU}. In comparison, the corresponding \texttt{PyMultiNest} retrieval requires $\sim49$ hours on 60 CPU cores, corresponding to approximately 2915 CPU-hours. Because the neural-network and Bayesian calculations use different hardware and algorithms, these values should not be interpreted as a direct hardware-normalized speed-up factor. The principal computational advantage instead arises from amortization: the training cost is incurred once, after which posterior inference can be performed rapidly for additional spectra that remain within the parameter and observational domain represented by the training distribution. \tr{The up-front training cost, model-specific training requirements, and need to assess posterior reliability nevertheless remain important considerations when extending the framework to new atmospheric models or observational configurations.} This scaling makes \texttt{NEXONN} particularly promising for large-sample atmospheric retrievals, repeated inference applications, population-level studies, and comparative exoplanetology in the JWST era.

\end{enumerate}

\section*{Acknowledgments}
\tr{We thank the anonymous referee for insightful comments and constructive suggestions that have significantly improved
the scientific rigor, clarity, and overall quality of this manuscript.} L.M. acknowledges funding support from the DAE through the NISER project RNI 4011. L.M. also gratefully acknowledges support from Breakthrough Listen at the University of Oxford through a sub-award at NISER under Agreement R82799/CN013, provided as part of a global collaboration under the Breakthrough Listen project funded by the Breakthrough Prize Foundation. The Open Access Fees for this article was fully funded by the One Nation One Subscription (ONOS) Initiative of Govt. of India (https://ror.org/036h6g940).
 
\vspace{5mm}
\facilities{JWST}
\software{NEXOTRANS \citep{Deka_2025}, PyTorch \citep{paszke2019pytorch}, PandExo \citep{pandexo_2017}, scikit-learn \citep{scikit-learn}, SciPy \citep{2020SciPy-NMeth}.}

\clearpage
\newpage

\bibliography{main}

\end{document}